\documentclass[12pt,authoryear,review]{elsarticle}

\makeatletter
\def\ps@pprintTitle{%
 \let\@oddhead\@empty
 \let\@evenhead\@empty
 \def\@oddfoot{}%
 \let\@evenfoot\@oddfoot}
\makeatother

\usepackage{tikz}
\usetikzlibrary{snakes}
\usetikzlibrary{positioning}
\usetikzlibrary{arrows.meta}
\usetikzlibrary{calc}
\usepackage{color}
\usepackage{booktabs}
\usepackage{mathtools}
\usepackage{amsmath}
\usepackage{amsthm}
\usepackage{amssymb}
\usepackage{amsfonts}
\usepackage{mathrsfs}
\usepackage{graphics}
\usepackage{graphicx}
\usepackage{dsfont}
\usepackage{subfigure}
\usepackage{url}
\usepackage[titletoc,title]{appendix}
\usepackage{enumitem}
\usepackage{rotating}
\usepackage{pdflscape}
\usepackage{afterpage}
\usepackage[normalem]{ulem}
\usepackage{caption}
\usepackage{float}
\usepackage{multirow}

\setitemize{itemsep=0pt}
\numberwithin{equation}{section}

\theoremstyle{plain}
\newtheorem{Theorem}{Theorem}[section] \newtheorem*{theorem*}{Theorem}
 \newtheorem*{proposition*}{Proposition}
 \newtheorem*{lemma*}{Lemma}
 \newtheorem*{corollary*}{Corollary}

\theoremstyle{definition}
 \newtheorem*{definition*}{Definition}
 \newtheorem*{example*}{Example}
\newtheorem{Remark}[Theorem]{Remark} \newtheorem*{remark*}{Remark}
 \newtheorem{Assumption}[Theorem]{Assumption}

\newcommand{\EE}{{\mathbb E}}

\newcommand{\RR}{{\mathbb R}}

\newcommand{\As}{{\mathcal A}}
\newcommand{\Ls}{{\mathscr L}}

\newcommand{\vX}{{\boldsymbol X}}
\newcommand{\vx}{{\boldsymbol x}}

\newcommand{\vb}{{\boldsymbol b}}
\newcommand{\vW}{{\boldsymbol W}}

\newcommand{\vu}{{\boldsymbol u}}
\newcommand{\vg}{{\boldsymbol g}}
\newcommand{\vw}{{\boldsymbol w}}
\newcommand{\vm}{{\boldsymbol m}}
\newcommand{\vy}{{\boldsymbol y}}
\newcommand{\vz}{{\boldsymbol z}}

\newcommand{\vtheta}{{\boldsymbol \theta}}

\newcommand{\Ito}{It\^{o} }

\newcommand\yuri[1]{{\color{black} #1}}

\definecolor{cadmiumgreen}{rgb}{0.0, 0.42, 0.24}
\definecolor{ceruleanblue}{rgb}{0.16, 0.32, 0.75}

\newcommand\redunderbrace[2]{{\color{ceruleanblue} \underbrace{\color{black} #1}_{\tiny \text{#2}} }}

\usepackage{array}
\newcolumntype{L}[1]{>{\raggedright\let\newline\\\arraybackslash\hspace{0pt}}m{#1}}
\newcolumntype{C}[1]{>{\centering\let\newline\\\arraybackslash\hspace{0pt}}m{#1}}
\newcolumntype{R}[1]{>{\raggedleft\let\newline\\\arraybackslash\hspace{0pt}}m{#1}}

\begin{document}

\begin{frontmatter}

\setlength{\parskip}{0em}

\title{\textbf{Extensions of the Deep Galerkin Method}
\\[1em]}

\author[1]{Ali Al-Aradi}
\author[2]{Adolfo Correia}
\author[3]{Gabriel Jardim}
\author[4]{Danilo de Freitas Naiff}
\author[5]{Yuri Saporito}

\address[1]{Department of Statistical Sciences, University of Toronto, Canada}
\address[2]{Instituto de Matem\'{a}tica Pura e Aplicada, Brazil}
\address[3]{Department of Statistics, Northwestern University, United States of America}
\address[4]{Instituto de Matem\'atica, Universidade Federal do Rio de Janeiro, Brazil}
\address[5]{Escola de Matem\'{a}tica Aplicada, Funda\c{c}\~ao Getulio Vargas, Brazil}

\begin{abstract}
We extend the Deep Galerkin Method (DGM) introduced in \cite{sirignano2018dgm} to solve a number of partial differential equations (PDEs) that arise in the context of optimal stochastic control and mean field games. First, we consider PDEs where the function is constrained to be positive and integrate to unity, as is the case with Fokker-Planck equations. Our approach involves reparameterizing the solution as the \textit{exponential} of a neural network appropriately normalized to ensure both requirements are satisfied. This then gives rise to nonlinear a partial integro-differential equation (PIDE) where the integral appearing in the equation is handled by a novel application of importance sampling. Secondly, we tackle a number of Hamilton-Jacobi-Bellman (HJB) equations that appear in stochastic optimal control problems. The key contribution is that these equations are approached in their \textit{unsimplified} primal form which includes an optimization problem as part of the equation. We extend the DGM algorithm to solve for the value function and the optimal control \textit{simultaneously} by characterizing both as deep neural networks. Training the networks is performed by taking alternating stochastic gradient descent steps for the two functions, a technique inspired by the policy improvement algorithms (PIA). 
\end{abstract}

\begin{keyword}
Partial differential equations;
Stochastic control;
Hamilton-Jacobi-Bellman equations;
Deep Galerkin Method;
Neural networks; 
Policy improvement.
\end{keyword}

\end{frontmatter}

\section{Introduction}

Partial differential equations (PDEs) are ubiquitous in many areas of science, engineering, economics and finance. They are often used to describe natural phenomena and model multidimensional dynamical systems. In the context of finance, finding solutions to PDEs is crucial for problems of derivative pricing, optimal investment, optimal execution, mean field games and many more. Although it is possible to obtain closed-form solutions to some PDEs, more often we must resort to numerical methods for arriving at an approximated solution. Traditional numerical approaches are presented in \cite{achdou2005computational}, \cite{brandimarte2013numerical} and \cite{burden2001numerical}. However, many of these classical approaches - particularly grid-based approaches such as finite difference methods - are burdened with issues of instability and computational cost, especially in higher dimensions. An alternative is to resort to Monte Carlo methods by appealing to the Feynman-Kac theorem to represent the solution to the PDE as an expectation and simulating to solve for the unknown function. This is primarily used for a class of linear PDEs although Monte Carlo methods for nonlinear PDEs have also been developed, e.g. \cite{gobet2005regression}. 

In recent years, a number of approaches utilizing techniques from machine learning have been developed to overcome the curse of dimensionality faced by mesh-based methods. These approaches often involve characterizing the unknown function using a deep neural network. For example, the work of \cite{weinan2017deep} and \cite{han2018solving} uses a \textit{deep BSDE method} which reformulates the nonlinear PDE of interest in terms of a backward stochastic differential equation (BSDE) by means of a nonlinear Feynman-Kac formula and then approximates the gradient of the unknown function by a neural network. An extension of this method is presented in \cite{beck2019deep} and \cite{hure2019some}. In addition, there are a number of numerical algorithms based on multilevel Picard iterations that provably overcome the curse of dimensionality for general nonlinear heat equations under Lipschitz assumptions, see e.g. \cite{e2016multilevel} and \cite{hutzenthaler2019overcoming}.

In contrast, the main idea behind solving PDEs using the Deep Galerkin Method (DGM) described in the work of \cite{sirignano2018dgm} is to represent the unknown function of interest using a deep neural network. Noting that the function must satisfy a known PDE, the network is trained by minimizing losses related to the differential operator acting on the function along with any initial, terminal and/or boundary conditions the solution must satisfy. The training data for the neural network consists of different possible inputs to the function and is obtained by sampling randomly from the region on which the PDE is defined. One of the key features of this approach is the fact that, unlike other commonly used numerical approaches such as finite difference methods, it is \textit{mesh-free}. Simulations indicate that the DGM may not suffer (as much as other numerical methods) from the curse of dimensionality associated with high-dimensional PDEs and PDE systems. A discussion of DGM and its applications can be found in \cite{alaradi2018solving}. On a related note, the work of \cite{hutzenthaler2019proof} proves that deep learning-based algorithms overcome the curse of dimensionality in the numerical approximation of solutions for a class of nonlinear PDEs.

\captionsetup[figure]{name={Fig}}

This paper addresses two perceived shortcomings of DGM. First, if the unknown function in the PDE is constrained in a certain way (for example if it is a probability density function that must be positive and integrate to unity), applying DGM does not guarantee that these constraints will be satisfied by the approximating neural network. This is true even when the constraints are directly incorporated into the loss function used to train the network. We propose a reparameterization that overcomes this difficulty.

The second issue is tied to Hamilton-Jacobi-Bellman (HJB) equations that arise in the context of stochastic control problems. Recall that such problems involve a controlled \Ito process $\vX^\vu = (\vX^\vu_t)_{t \geq 0}$ satisfying the stochastic differential equation

\[ 
d\vX^\vu_t = \mu(t, \vX_t^\vu, \vu_t) \ dt + \sigma(t, \vX_t^\vu, \vu_t)  \ d\vW_t \ , \qquad \qquad \vX_0^\vu = \vx_0 \, , 
\]  
where $\vu = (\vu_t)_{t\geq0}$ is a control process chosen by the controller from an admissible set $\mathbb{A}$ taking values in $\As$. For a given control, the agent's performance criteria is:
\begin{align*}
H^\vu(t,\vx) &= \EE \left[ \int_{t}^{T} F(s,\vX_s^\vu,\vu_s) ~ds + G(\vX_T^\vu) ~\middle|~ \vX^\vu_t = \vx \right].
\end{align*} \\
Assuming enough regularity, the value function $H(t,\vx) = \underset{\vu \in \mathbb{A}}{\sup} ~ H^\vu(t,\vx)$, can be shown to satisfy a fully nonlinear PDE referred to as the HJB equation given by
\begin{equation*}
\begin{cases}
\partial_t H(t,\vx) +  \underset{\vu \in \As}{\sup}  ~ \left\{ \Ls^\vu_t H(t,\vx) + F(t,\vx,\vu) \right\}  = 0,
\\
H(T,\vx) = G(\vx),
\end{cases}
\end{equation*} \\
where the differential operator $\Ls^\vu_t$ is the infinitesimal generator of the controlled process $\vX^\vu$. We will refer to this unsimplified form of the HJB equation as the \textit{primal form}.

It is sometimes possible to simplify the primal form of an HJB equation by analytically solving for the optimal control in feedback form (i.e. expressed in terms of the value function and its derivatives) and substituting this quantity back into the HJB equation. This removes the optimization step that appears as the second term of the HJB equation and leaves us with a more familiar form for the PDE that can be handled well by DGM. In fact, \cite{alaradi2018solving} and \cite{sirignano2018dgm} both demonstrate the application of DGM to HJB equations simplified in this manner. However, sometimes it is not possible to arrive at such a simplification, and in those cases DGM would not be able to handle the optimization step. Furthermore, even in situations where the primal form can be simplified and DGM can be successfully applied to approximate the value function, we are still left with translating the value function to the optimal control. We find that the error propagation in this step can lead to unsatisfactory results for the optimal control, which is arguably the main object of interest. Instead, our approach addresses both of these issues by parameterizing the unknown value function as well as the unknown optimal control as deep neural networks and training the two networks by taking alternating stochastic gradient descent steps. This is similar in spirit to the approach used in policy improvement algorithms commonly employed in reinforcement learning problems.

\yuri{For the numerical examples shown below, we have chosen PDE problems with available closed-form solution in order to compare the numerical solution with the true one and to quantify the error generated by the numerical method. Moreover, it gives the reader a controlled setting where every ingredient of the problem is well understood.}

The remainder of this article is organized as follows: we conclude this section by describing the implementation details of DGM. Section \ref{sec:FKequations} tackles the problem of PDEs with integration and positivity constraints using the Fokker-Planck equation as an example. In Section \ref{sec:HJBequations}, we present a modified DGM algorithm, called DGM-PIA, and apply it to solving three HJB equations, namely the Merton problem of optimal investment, an optimal execution problem and the multidimensional Linear-Quadratic problem. In Section \ref{sec:HJBsystems} we apply the DGM algorithm to a stochastic game which involves multiple agents leading to a system of HJB equations. Finally, in Section \ref{sec:meanFieldGames} we apply the DGM algorithm combined with the technique discussed in Section \ref{sec:FKequations} to solve a mean-field game version of the optimal execution problem.

\subsection{Implementation Details}

The architecture adopted by \cite{sirignano2018dgm} is similar to that of LSTMs and Highway Networks described in \cite{hochreiter1997long} and \cite{srivastava2015highway}, respectively. It consists of three layers, which we refer to as \textbf{DGM layers}: an input layer, a hidden layer and an output layer, though this can be easily extended to allow for additional hidden layers. \\

\begin{figure}[h!]
	\centering
	\scalebox{0.7}{
	\begin{tikzpicture}
	
	\definecolor{yellowd}{RGB}{255,187,51}
	\definecolor{yellowf}{RGB}{255,221,153}
	\definecolor{greend}{RGB}{102,255,102}
	\definecolor{greenf}{RGB}{153,255,153}
	\definecolor{blued}{RGB}{51,187,255}
	\definecolor{bluef}{RGB}{153,221,255}
	\definecolor{pinkd}{RGB}{210,121,164}
	\definecolor{pinkf}{RGB}{230,179,204}
	
	\tikzstyle{scircle}=[circle, thick, draw=greend, fill=greenf, minimum size=1cm]
	\tikzstyle{xcircle}=[circle, thick, draw=blued, fill=bluef, minimum size=1cm]
	\tikzstyle{ycircle}=[circle, thick, draw=pinkd, fill=pinkf, minimum size=1cm]
	\tikzstyle{smallrect}=[rectangle, thick, draw=yellowd, fill=yellowf, minimum height=0.75cm, minimum width=1.5cm,]
	\tikzstyle{dgmlayer}=[rectangle, thick, draw=yellowd, fill=yellowf, minimum height=3cm, minimum width=1.5cm, rounded corners]
	\tikzstyle{arrow}=[draw=black, -{Latex[length=2.5mm]}, thick]

	\node[smallrect] (R1) {$\vw^1 \cdot \vx + \vb^1$};
	\node[scircle, below=1.0cm of R1] (S1) {$S^1$};
	\node[xcircle, above=1.0cm of R1] (X) {$\vx$};
	
	\node[dgmlayer, right=1.5cm of S1, yshift=1cm] (L1) {};
	\node[rotate=-90] at (L1) {DGM Layer};
	\node[dgmlayer, right=1.0cm of L1] (L2) {};
	\node[rotate=-90] at (L2) {DGM Layer};
	\node[dgmlayer, right=1.0cm of L2] (L3) {};
	\node[rotate=-90] at (L3) {DGM Layer};
	
	\node[scircle, right=1.0cm of L3, yshift=-1cm)] (SL) {$S^{L+1}$};
	\node[smallrect, right=of SL] (RL) {$\vw \cdot S^{L+1} + \vb$};
	\node[ycircle, right=of RL] (Y) {$\vy$};
	
	\draw[arrow] (X) -- (R1);
	\draw[arrow] (R1) -- node[anchor=east] {$\sigma$} (S1);
	\draw[arrow] (S1) -- (S1-|L1.west);
	\draw[arrow] (S1-|L1.east) -- (S1-|L2.west);
	\draw[arrow] (S1-|L2.east) -- (S1-|L3.west);
	
	\draw[arrow] (X) -| coordinate (aL1) (L1);
	\draw[arrow] (aL1) -| coordinate (aL2) (L2);
	\draw[arrow] (aL2) -| (L3);
	
	\draw[arrow] (S1-|L3.east) -- (SL);
	\draw[arrow] (SL) -- (RL);
	\draw[arrow] (RL) -- (Y);

	\end{tikzpicture}
	}
	\captionsetup{width=.9\linewidth}
	\caption{Bird's-eye perspective of overall DGM architecture.}
	\label{fig:networkoutside}
\end{figure}

From a bird's-eye perspective, each DGM layer takes as an input the original mini-batch inputs $\vx$ (in our case this is the set of randomly sampled time-space points) and the output of the previous DGM layer. This process culminates with a vector-valued output $\vy$ which consists of the neural network approximation of the desired function $u$ evaluated at the mini-batch points. See Figure \ref{fig:networkoutside} for a visualization of the overall architecture. \\

Within a DGM layer, the mini-batch inputs along with the output of the previous layer are transformed through a series of operations that closely resemble those in Highway Networks. Below, we present the architecture in a visual representation of a single DGM layer in Figure \ref{fig:dgmlayer3}:

\begin{figure}[h!]
	\centering
	\scalebox{0.7}{
	\begin{tikzpicture}
	
	\definecolor{yellowd}{RGB}{255,187,51}
	\definecolor{yellowf}{RGB}{255,221,153}
	\definecolor{greend}{RGB}{102,255,102}
	\definecolor{greenf}{RGB}{153,255,153}
	\definecolor{blued}{RGB}{51,187,255}
	\definecolor{bluef}{RGB}{153,221,255}
	\definecolor{pinkd}{RGB}{210,121,164}
	\definecolor{pinkf}{RGB}{230,179,204}
	
	\tikzstyle{scircle}=[circle, thick, draw=greend, fill=greenf, minimum size=1cm]
	\tikzstyle{xcircle}=[circle, thick, draw=blued, fill=bluef, minimum size=1cm]
	\tikzstyle{ycircle}=[circle, thick, draw=pinkd, fill=pinkf, minimum size=1cm]
	\tikzstyle{smallrect}=[rectangle, thick, draw=yellowd, fill=yellowf, minimum height=1.5cm, minimum width=4.0cm, rounded corners]
	\tikzstyle{arrow}=[draw=black, -{Latex[length=2.5mm]}, thick]
	
	\node[] (SX1) {};
	\node[right=of SX1] (SX2) {};
	
	\node[xcircle, left=of SX1, yshift=1cm] (Sold) {$S$};
	\node[xcircle, left=of SX1, yshift=-1cm] (X) {$\vx$};
	
	\node[smallrect, right=of SX2] (GR) {$\vu^z \cdot \vx + \vw^z \cdot S + \vb^z$};
	\node[smallrect, above=of GR] (ZR) {$\vu^g \cdot \vx + \vw^g \cdot S + \vb^g$};
	\node[smallrect, below=of GR] (RR) {$\vu^r \cdot \vx + \vw^r \cdot S + \vb^h$};
	
	\node[scircle, right=of ZR] (Z) {$Z$};
	\node[scircle, right=of GR] (G) {$G$};
	\node[scircle, right=of RR] (R) {$R$};
	
	\node[smallrect, right= 1.45cm of Z] (SR) {$(1-G) \odot H + Z \odot S$};
	\node[smallrect, right=of R] (HR) {$\vu^h \cdot \vx + \vw^h \cdot (S \odot R) + \vb^h$};
	\node[scircle] (H) at ($(HR)!0.5!(SR)$) {$H$};
	\node[ycircle, right=of SR] (Snew) {$S^{new}$};

	\draw[thick] (Sold) -| (SX1.center);
	\draw[thick] (X) -| (SX1.center);
	\draw[arrow] (SX1.center) -- (GR);
	\draw[arrow] (SX2.center) |- (ZR);
	\draw[arrow] (SX2.center) |- coordinate (lRR) (RR);
	\draw[arrow] (ZR) -- node[above] {$\sigma$} (Z);
	\draw[arrow] (GR) -- node[above] {$\sigma$} (G);
	\draw[arrow] (RR) -- node[above] {$\sigma$} (R);
	\draw[arrow] (Z) -- (SR);
	\draw[arrow] (R) -- (HR);
	\draw[arrow] (HR) -- node[left] {$\sigma$} (H);
	\draw[arrow] (H) -- (SR);
	\draw[arrow] (G) -| ([xshift=-1.0cm]SR.south);
	\draw[arrow] (SR) -- (Snew);
	
	\node[above=of ZR] (aZR) {};
	\draw[thick] (Sold) |- (aZR.center);
	\draw[arrow] (aZR.center) -| (SR);
	\node[below=of RR] (bRR) {};
	\draw[thick] (lRR) |- (bRR.center);
	\draw[arrow] (bRR.center) -| (HR);

	\end{tikzpicture}
	}
	\captionsetup{width=.9\linewidth}
	\caption{Operations within a single DGM layer.  Here $\odot$ denotes Hadamard (element-wise) multiplication, $\sigma$ is an activation function and the $\vu$, $\vw$ and $\vb$ terms with various superscripts are the model parameters.}
	\label{fig:dgmlayer3}
\end{figure}
Similar to the intuition for LSTMs, each layer produces weights based on the last layer, determining how much of the information gets passed to the next layer. In \cite{sirignano2018dgm} the authors also argue that including repeated element-wise multiplication of nonlinear functions helps capture ``sharp turn'' features present in more complicated functions. 

Compared to a Multilayer Perceptron (MLP), the number of parameters in each hidden layer of the DGM network is roughly eight times bigger than the same number in a usual dense layer. This is the case because each DGM layer has 8 weight matrices and 4 bias vectors while the MLP network only has one weight matrix and one bias vector (assuming the matrix/vector sizes are similar to each other). Moreover, the LSTM-like architecture of DGM networks is able to handle issues of vanishing gradients - an issue that deep MLPs may encounter - while being flexible enough to model complex functions. Note that at every iteration the original input enters into the calculations of every intermediate step, thus decreasing the chance of vanishing gradients of the output function with respect to $\vx$.

As noted by \cite{sirignano2018dgm}, the architecture of a neural network can be crucial to its success and clever choices of architectures, which exploit a priori knowledge about an application, can significantly improve performance. However, in our work we maintain the same architecture for the neural network, but may slightly modify the parametrization of the approximating function.

In most examples below, we consider $L=3$ layers each one with $n=64$ neurons. The batch size $M$ depends on the particular PDE, but it is generally chosen around $2^{10}$. Moreover, unless stated otherwise, we consider $N=50,000$ iterations and the update of the network's parameters by the Adaptive Moment Estimation (Adam) for ten epochs with mini-batch equals $M$. Weights and biases are initialized using the Glorot uniform initializer. Additionally, we have used either constant learning rate or the schedule presented in \cite{sirignano2018dgm}.

\section{PDEs with Integration and Positivity Constraints} \label{sec:FKequations}

\setcounter{figure}{0}

\subsection{Fokker-Planck Equations}\label{sec:FK_method}

In this section we tackle the problem of applying DGM when the unknown function in the PDE is constrained to be positive and integrate to unity. As an example, this is the case when we are interested in solving a Fokker-Planck equation to obtain the time evolution of a probability density function associated with a diffusion process of interest. In particular, assume that $\vX = (\vX_t)_{t \geq 0}$ is an \Ito process on $\RR^d$ satisfying the stochastic differential equation (SDE)
\begin{align}\label{eq:sde}
d\vX_t = \mu(\vX_t) \, dt + \sigma(\vX_t) \, d\vW_t \, ,
\end{align}
where $\vW$ is a $k$-dimensional standard Brownian motion and the initial point is a random vector $\vX_0$ with distribution given by a probability density function $f$. Let us consider the following regularity assumption on $\mu$ and $\sigma$:
\begin{Assumption}\label{assump:fokker_planck}
The coefficients $\mu$ and $\sigma$ are smooth with bounded derivatives of all orders and the matrix $a=\sigma \sigma^\intercal$ is uniformly elliptic, i.e. there exists $\alpha > 0$ such that $\sum_{i,j} a_{ij}(\vx) \xi_i \xi_j \geq \alpha \|\boldsymbol{\xi}\|^2$, for all $\vx, \boldsymbol{\xi} \in \RR^d$.
\end{Assumption}
Under the assumption above, the random vector $\vX_t$ has a smooth probability density function, denoted by $p(t,\vx)$, satisfying the PDE:
{ \small 
\begin{equation}\label{eq:fokker_planck}
\begin{cases}
\displaystyle \partial_t p + \sum_{j=1}^{d} \partial_j (\mu_j(\vx) \, p) - \frac{1}{2}  \sum_{i,j=1}^{d} \partial_{ij} (a_{ij}(\vx) \, p) = 0, ~~~ (t,\vx) \in \RR_+ \times \RR^d  \, ,
\\
p(0,\vx) = f(\vx), ~~~ \vx \in \RR^d  \, ,
\end{cases}
\end{equation} 
}%
see, for instance, \cite{book_fokker_planck}. Clearly, since $p(t,\cdot)$ is a probability density function we should have $p(t,\vx) \geq 0$, for all $(t,\vx)$, and $\int_{\RR^d} p(t,\vx) \, d\vx = 1$, for all $t$. 

\begin{figure}
\centering
\includegraphics[width=\linewidth, trim={0 1cm 0 1cm}, clip]{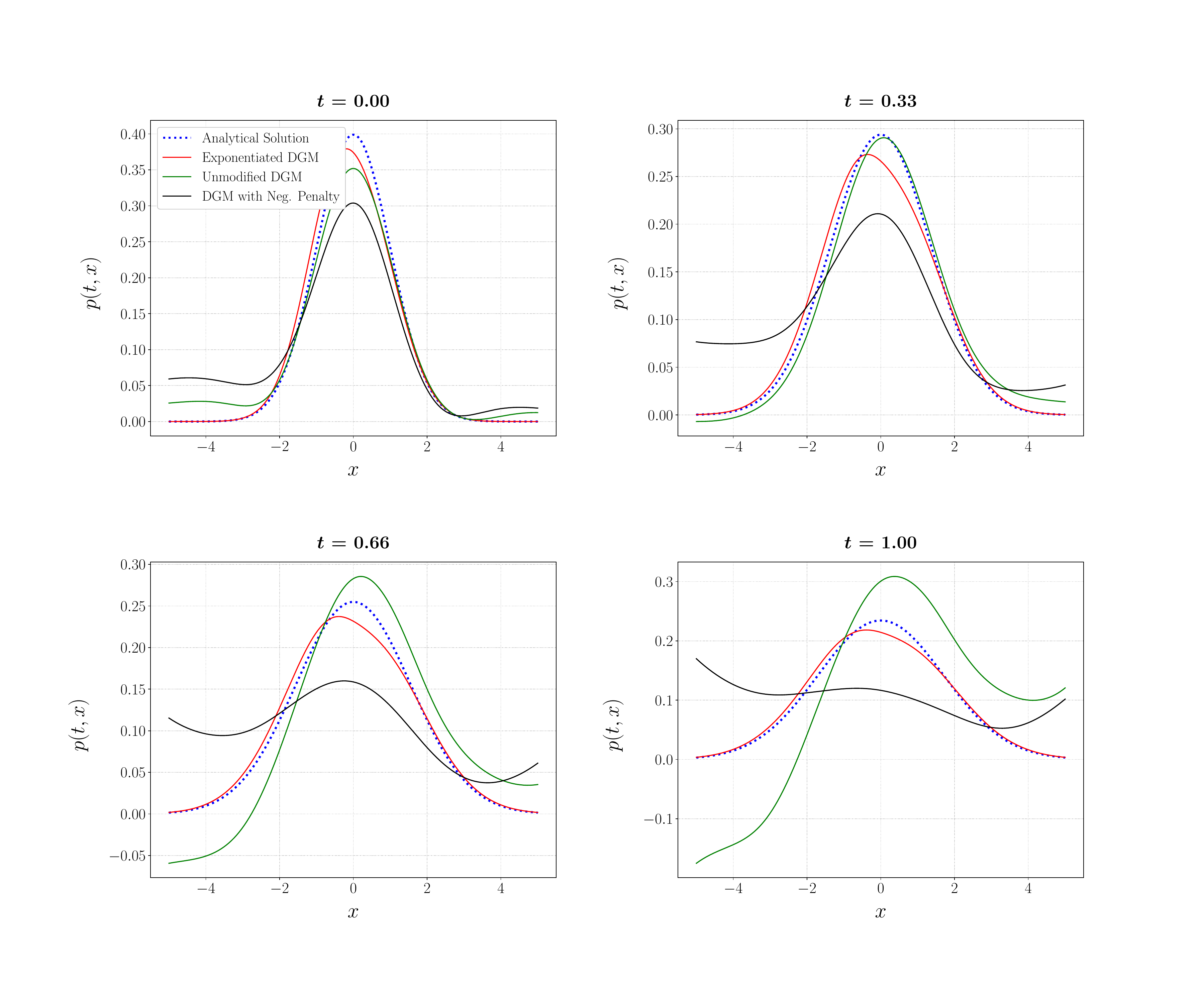}
	\caption{Distribution of OU process $X_t$ with random Gaussian starting point at different times approximated using DGM. The parameters chosen are $d=1$, $B = 0.5$, $m = 0$, $C=2$, $T=1$, $m_0 = 0$ and $C_0 = 1$. We have run the algorithm for $N=500$ iterations.}
	\label{fig:fokkerPlanck_density}
\end{figure}

However, directly applying the DGM algorithm to solve the Fokker-Planck equation does not guarantee that the positivity and integration constraints will be satisfied. This is true even when the constraints are directly incorporated into the loss function used to train the network. To demonstrate this we apply the DGM algorithm on the Fokker-Planck equation for the one-dimensional Ornstein-Uhlenbeck (OU) process with a random Gaussian starting point. We add an additional term in the loss function (besides the usual terms for the differential operator and the initial condition) to reflect the non-negativity constraint, namely:
\[
L_{pos}(\vtheta_n;t_n,x_n) = \max \{-f(t_n,x_n; \vtheta_n), 0\}.
\] 
We initially included a penalty term to force the integral of the approximating function to equal one, however this proved to be too computationally expensive due to the fact that a numerical integration procedure had to run at each step of the network training phase. Instead, we opted for normalizing the density function after the estimation was completed. Figure \ref{fig:fokkerPlanck_density} shows the results of this approach for the one-dimensional example. The plots show that, while the initial condition was somewhat well-approximated, the fitted distributions using the unmodified and penalized versions of the DGM had issues around the tails of the distribution at later times and that the Gaussian bell shape was not conserved across time. This discrepancy diminishes when we run the algorithm for a significant larger number of iterations, but the behavior, specially on the tails, are still unsatisfactory and accentuated for large $t$. 

To address these issues while avoiding the computational difficulties that come with incorporating the integration penalty directly, we first reparameterize the problem by writing the density function as a normalized exponentiated function. This allows us to derive a related nonlinear partial integral differential equation (PIDE) that automatically incorporates both constraints. 

\vspace{2cm}
\begin{Theorem} \label{thm:reparam}
Let $\vX = (\vX_t)_{t \geq 0}$ be the \Ito process (\ref{eq:sde}) satisfying Assumption \ref{assump:fokker_planck} and $p$ its probability density. Let $u$ be any solution\footnote{See Remark \ref{rmk:non_uniqueness} for a discussion on the non-uniqueness of solutions for the PDE (\ref{eq:pde_u}).} of the PIDE below
{\small 
\begin{equation}\label{eq:pde_u}
\begin{cases}
~  \displaystyle \partial_t u - \frac{\int_{\RR^d} e^{-u(t,\vy)  } \partial_t u(t,\vy) \, d\vy}{\int_{\RR^d} e^{-u(t,\vy)} \, d\vy} + \sum_{j=1}^d \mu_j \, \partial_j u - \partial_j \mu_j   \\
\displaystyle - \tfrac{1}{2} \sum_{i,j=1}^d \left[ -\partial_{ij} a_{ij} + \partial_i a_{ij} \, \partial_j u + \partial_j a_{ij} \, \partial_i \, u + a_{ij} \Big(\partial_{ij}u - \partial_j u \, \partial_i u \Big) \right]   = 0 \, ,
\\[2em]
~ u(0,\vx) = -\log( \tilde{f}(\vx)),
\end{cases}
\end{equation}}%
where $\tilde{f}$ is any positive function proportional to $f$. Then
\begin{align}\label{eq:p}
p\left(t,\vx\right)=\frac{e^{-u\left(t,\vx\right)}}{\int_{\RR^d} e^{-u(t,\vy)} \, d\vy} \, . 
\end{align}
is the unique solution to the PDE \eqref{eq:fokker_planck}.
\end{Theorem}
\begin{proof}
See \ref{sec:appendix_reparam}.
\end{proof}

\begin{Remark}\label{rmk:non_uniqueness}
Notice that if $u$ is a solution to PDE (\ref{eq:pde_u}), then so is $u(t,\vx) + h(t)$, for any $h$ differentiable with $h(0) = 0$. Moreover, notice the integral term does not depend on $\vx$ and then the PDE that $\partial_k u$ satisfies does not have a integral term, becoming a well-studied non-linear PDE that, under mild regularity assumptions on its coefficients, has a unique solution. Therefore, we are able to conclude that all classical solutions of (\ref{eq:pde_u}) is of the form $u(t,\vx) + h(t)$. However, all these solutions yield the same $p$ from Equation (\ref{eq:p}).
\end{Remark}

This new Equation (\ref{eq:pde_u}) is a non-linear partial integro-differential equation (PIDE). To handle the integral term and avoid the costly operation of numerically integrating at each step, we notice that since we have uniformly sampled the mini-batch $\{t_j\}_{j=1}^{N_t}$ from $[0,T]$ and $\{\vx_k\}_{q=1}^{N_x}$ from $\mathbb{R}^d$ at each iteration as part of the DGM algorithm, we can use \textbf{importance sampling} to approximate the integral for each $t_j$. That is,
\[ \int_{\RR^d} \partial_t u (t_j,\vx) ~ \frac{ e^{-u(t_j,\vx)  } }{\int_{\RR^d} e^{-u(t_j,\vy)}d\vy} ~ d\vx ~\approx~ \sum_{k=1}^{N_x} \partial_t u(t_j,\vx_k) \ w(\vx_k) \,, \] 
where 
\[ w(\vx) = \frac{e^{-u(t_j,\vx)}}{\sum_{i=1}^{N_x} e^{-u(t_j,\vx_i)} } \, . \] 

Note that this procedure can be adapted to other PDEs with similar constraints. Applying the DGM algorithm to this PDE to solve for $u(t,\vx)$, integrating $e^{-u}$ and then translating the output back to the density function $p(t,\vx)$ as in Equation (\ref{eq:p}) guarantees that the resulting approximation will remain positive and integrate to unity.

\subsection{Application to multidimensional Ornstein-Uhlenbeck processes}\label{sec:OU}

Let $\vX = (\vX_t)_{t\geq 0}$ be a multidimensional Ornstein-Uhlenbeck (OU) process:
\[ d\vX_t = (\vm - B\vX_t ) ~dt + C \, d\vW_t. ,\]
Assume further that the process starts at a random point drawn from a normal distribution with mean $\vm_0$ and covariance $C_0$, i.e. $\vX_0 \sim \mathcal{N}(\vm_0,C_0)$. It is straightforward to show that the distribution of $\vX_t$ is  normal:
\[
\vX_{t} ~\sim~ \mathcal{N} \Bigg( \redunderbrace{e^{-Bt}(\vm_0 - \vm) + \vm }{$\vm(t)$}, ~ \redunderbrace{e^{-Bt} C_0 e^{-B^\intercal t} + \int_0^t e^{-B(t-s)} CC^\intercal e^{-B^\intercal(t-s)}ds}{$C(t)$} \Bigg) \, ,
\]
where $B^\intercal$ is the transpose of $B$. From the distribution of $\vX_t$ given above we can deduce that the solution to this PDE is
\[
p(t,\vx) = \frac{1}{\sqrt{(2 \pi)^d |C(t)|} } ~ \exp \left[ -\frac{1}{2} (\vx - \vm(t))^\intercal C(t)^{-1} (\vx - \vm(t)) \right]\, ,
\]
where $|C|$ is the determinant of $C$. We can use this result to assess the accuracy of our numerical solutions to the Fokker-Planck PDE.

Using the alternative approach, we apply Theorem \ref{thm:reparam} to obtain the PDE that the function $u$ should satisfy, which is
\begin{align}\label{eq:pde_u_OU}
\begin{cases}
~  \partial_t u + \nabla u^\intercal (\vm - B\vx) + Tr(B) - \tfrac{1}{2} Tr(A \ \tilde{H}u)  -\displaystyle\frac{\int_{\RR^d} e^{-u(t,\vy)  } \partial_t u(t,\vy) \, d\vy}{\int_{\RR^d} e^{-u(t,\vy)} \, d\vy} = 0,
\\\\
~  u(0,\vx) = \frac{1}{2}(\vx - \vm_0)^\intercal C_0^{-1} (\vx - \vm_0),
\end{cases}
\end{align}
where $A = CC^\intercal$ and $(\tilde{H}u)_{ij} = \partial_{ij} u - \partial_i u \partial_j u$. The results of the modified approach based on reparametrization are also given in Figure \ref{fig:fokkerPlanck_density} and the plots show a marked improvement over the alternative implementations of DGM to this problem. 

\begin{figure}
\begin{minipage}[b]{0.49\textwidth}
\centering
\includegraphics[width=\linewidth]{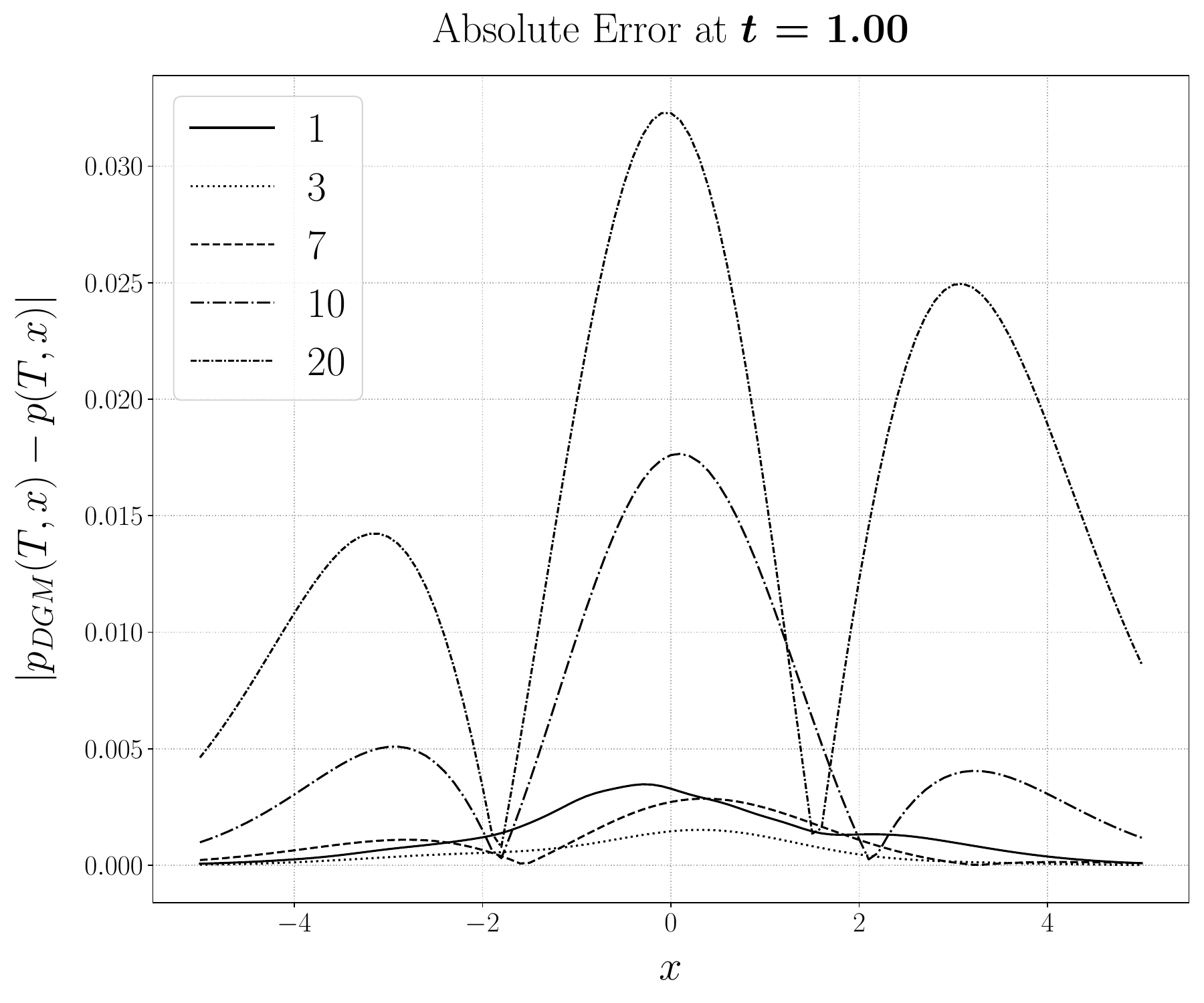}
\vspace{-1.2cm}
\caption{\yuri{Absolute error of the marginal density of the first dimension at time $t=1$.}} \label{fig:fokkerPlanck_density_dimension}
\end{minipage}
\hfill
\captionsetup[figure]{name={Table}}
\begin{minipage}[b]{0.49\textwidth}
\centering
\begin{tabular}{lcc}
\toprule
$d$  & MSE  & $p_{DGM}(1,m_0)$ \\
\midrule
1  &   0.000812 & 2.3770e-01  \\
3  &   0.000863 & 2.3491e-01  \\
7  &   0.010850 & 2.3371e-01 \\
10 &   0.017106 & 2.3019e-01  \\
20 &   0.037591 & 2.0214e-01 \\
\bottomrule
\end{tabular}
\captionof{table}{\yuri{The estimation of mean squared error between the approximation and the true marginal density of the first dimension. We also show the pointwise approximation at $m_0$; the true value is $p(1,m_0)=$ 2.3441e-01. For the approximation of the MSE, we used 101 points in the $x_1$ dimension between $-5$ and $5$ and 11 points in the time dimension between 0 and 1.}} \label{tab:fokkerPlanck_density_dimension}
\end{minipage}
\end{figure}

In order to exemplify the capabilities of the method for higher dimensional problems, we consider PDE (\ref{eq:pde_u_OU}) with $d \in \{1,3,7,10\}$. We have used the same architecture in all examples and considered the parameters chosen as $B = 0.5 I_d$, $m = 0$, $C=2I_d$, $T=1$, $m_0 = 0$ and $C_0 = I_d$. We ran the algorithm until a total loss of order $10^{-4}$ was attained or iteration of $N=50,000$. In order to access the accuracy of the algorithm, we computed the absolute mean-squared error (MSE) of a given marginal density.\footnote{For instance, the marginal density of the first variable $x_1$ is given by $p(t,x_1) = \int_{\RR^{d-1}} p(t,x_1,x_2,\ldots,x_d) dx_2\cdots dx_n$, where this integral is approximated numerically.} This avoids the issue that density functions defined in high-dimensional spaces usually take very small values because of the integral restriction. Moreover, relative errors should be avoided since the true marginal density take values too close to zero. From Table \ref{tab:fokkerPlanck_density_dimension}, we conclude that the algorithm is capable of dealing with increasingly higher dimensions PDEs even without escalating complexity of the DGM network. Using more computer power, the algorithm proposed here should be able to scale to higher dimension as shown in \cite{sirignano2018dgm} for the original DGM method.

\section{Hamilton-Jacobi-Bellman Equations} \label{sec:HJBequations}

\setcounter{figure}{0}

In this section we consider applying a modified version of DGM to solving HJB equations in their primal form, i.e. in the form 
\begin{equation}\label{eq:hjb}
\begin{cases}
\partial_t H(t,\vx) +  \redunderbrace{\underset{\vu \in \As}{\sup}  ~ \left\{ \Ls^{\vu} H(t,\vx) + F(t,\vx,\vu) \right\}}{optimization}  = 0,
\\
H(T,\vx) = G(\vx).
\end{cases}
\end{equation}
In the PDE above we are interested in solving for the unknown value function $H$ and the unknown optimal feedback control $\vu^*$ both of which are defined on the region $[0,T] \times \Omega$ where $\Omega \subset \RR^d$. As discussed in the outset, it would be difficult to apply DGM directly to an HJB equation of this nature due to the optimization component that appears in the PDE. The typical approach for a na\"{i}ve application of DGM would be to solve for the optimal control in feedback form, substitute it back into the PDE and eliminate the optimization component, making it possible to apply DGM directly. However, such a simplification is not always feasible. Moreover, solving for the value function using DGM and then translating the output to obtain the optimal control may lead to unsatisfactory results. This is due to possible instabilities that arise from the dependence of the optimal control on the derivatives of the value function.

The alternative we propose to overcome these difficulties is a modification of the DGM algorithm inspired by the policy improvement algorithm (PIA) used in reinforcement learning which allows us to obtain a numerical solution for both the value function and the optimal control simultaneously. 

We begin by presenting a brief summary of PIA. Given a control in feedback form $u$, denote $\Ls^u H (t,\vx) = \Ls^{u(t,\vx)} H(t,\vx)$ and let $u_0$ be an initial control. Then for $n \geq 0$ the algorithm involves alternating between two steps: 
\begin{enumerate}
\item Find (or approximate) a classical solution to the linear PDE
\begin{equation} \label{eq:linear_PDE}
\begin{cases}
\partial_t H^{\vu_n}(t,\vx) + \Ls^{\vu_n} H^{\vu_n}(t,\vx) + F(t,\vx,\vu_n(t,\vx)) = 0,
\\
H^{u_n}(T,\vx) = G(\vx),
\end{cases}
\end{equation}
for the fixed control $\vu_n(t,\vx)$.
\item Compute the policy improvement 
\begin{equation} \label{eq:PIA}
\vu_{n+1}(t,\vx) \in \underset{\vu \in \As}{\arg \max} ~ \Big\{ \Ls^{\vu} H^{\vu_n}(t,\vx) + F(t,\vx,\vu) \Big\}  
\end{equation} 
for the fixed value function $H^{\vu_n}$.
\end{enumerate}

The convergence of PIA has been studied in several frameworks, e.g. \cite{jacka2017pia} and other references therein.

\subsection{Modified Deep Galerkin Method for HJB Equations (DGM-PIA)}\label{sec:DGM_PIA}

The main idea of the algorithm proposed here is to combine the DGM and PIA algorithms. The modification to the DGM approach involves approximating the value function $H$ and the optimal control $\vu^*$ with functions $f(t,\vx;\vtheta^H)$ and $\vg(t,\vx;\vtheta^u)$ given by two deep neural networks with parameter sets $\vtheta^H$ and $\vtheta^u$. Since there are two optimization problems - one corresponding to the first-order equation, the other to the PDE satisfied by the value function - we define two loss functionals associated with training $f$ and $\vg$. The first loss functional addresses the differential operator and the terminal condition related to the PDE of the value function:
{\small \begin{align*} 
	L_H(\vtheta^H) &= \redunderbrace{\Big\| \left( \partial_t  + \Ls^{\vg(\cdot; \vtheta^u)} \right) f \left( \cdot; \vtheta^H \right) + F\big( \cdot,\vg(\cdot;\vtheta^u) \big) \Big\|_{[0,T] \times \Omega, \nu_1}^2}{\mbox{differential operator}} + 
	\redunderbrace{\Big\|f \left( T,\cdot; \vtheta^H \right) - G\Big\|_{\Omega, \nu_2}^2}{\mbox{terminal condition}}.
	\end{align*} 
}%
The second is associated with the auxiliary optimization problem:
\begin{equation*} 
	L_u(\vtheta^u) = \redunderbrace{-\int_{[0,T] \times \Omega} \Big[ \Ls^{\vg(t,\vx; \vtheta^u)} f \left( t,\vx; \vtheta^H \right) + F\big( t,\vx,\vg(t,\vx;\vtheta^u) \big) \Big] d\nu_1(t, \vx)}{optimization}.
	\end{equation*} 

This latter term is related to the policy improvement Equation \eqref{eq:PIA}, however it addresses this optimization in an average sense over the domain and sampled points rather than pointswise as in \eqref{eq:PIA}. In order to minimize the two loss functionals, we apply stochastic gradient descent in an alternating manner. That is, we take one Adam step for $\vtheta^H$, then fixing this parameter set value take one Adam step for $\vtheta^u$. The modified DGM algorithm is defined in detail in Algorithm \ref{fig:modifiedDGM}. The description given in Algorithm \ref{fig:modifiedDGM} should be thought of as a general outline that can be modified according to the particular nature of the HJB problem being considered. The number of iterations, $N$, of the DGM-PIA algorithm is run until a total loss $L_H$ attains a desired value and additional training does not decrease it.

\yuri{In contrast to the class of PDEs studied in the previous section, the problem we consider here is a terminal value problem, i.e. the boundary condition specifies the value function at the terminal time point T. In our algorithm and analysis, we consider the unknown function to be one of forward time $t$ and as a consequence we incorporate a penalty term based on the terminal condition in the loss function. Obviously, one could consider an approach where the flow of time is reversed, i.e. the function is defined in terms of backward time $\tau=T-t$ along with the appropriate initial condition included in the loss function.}

\captionsetup[figure]{name={Algorithm}}

\begin{figure}[h!]
	{\small
		\begin{tabular}{r}
			\hline\hline
			\hspace{0.95\textwidth}
		\end{tabular}
		\vspace{-0.5em}
		\begin{enumerate}
			\item Choose a loss tolerance $\varepsilon > 0$ and a maximum number of iterations $N$. Initialize parameter sets for the value function $\vtheta_0^H$ and optimal control $\vtheta_0^u$ and the associated learning rates $\alpha_n^H$ and $\alpha_n^u$.  
			\item Generate random samples from the domain's interior and \yuri{final condition}, i.e.
			\begin{itemize}
				\item \texttt{Generate $\{(t_m,\vx_m)\}_{m=1}^M$ from $[0,T] \times \Omega $ according to $\nu_1$} 
				\item \texttt{Generate \yuri{$\{ \vz_m\}_{m=1}^M$ from $\Omega$ according to $\nu_2$}}
			\end{itemize} 
			\item Compute the value function loss functional for the current mini-batch, i.e. the randomly sampled points $s_n = \left\{(t_m, \vx_m), \vz_m\right\}_{m=1}^M$: 
			\begin{itemize}
				\item $L_{H,1}(\vtheta_n^H;\{(t_m,\vx_m)\}_{m=1}^M) = \tfrac{1}{M}\sum_{m=1}^M\Big[ \Big( \partial_t  + \Ls^{\vg(t_m, \vx_m; \vtheta_n^u)} \Big) f \Big( t_m,\vx_m; \vtheta_n^H \Big)$ \\$+ F\big( t_m,\vx_m,\vg(t_m,\vx_m;\vtheta_n^u) \big) \Big]^2$ 
				\item \yuri{$L_{H,2}(\vtheta_n^H;\{\vz_m\}_{m=1}^M) = \tfrac{1}{M} \sum_{m=1}^M(f(T,\vz_m; \vtheta_n^H ) - G(\vz_m))^2 $}
				\item $L_H(\vtheta_n^H;s_n) = L_{H,1}(\vtheta_n^H;\{(t_m,\vx_m)\}_{m=1}^M)) + L_{H,2}(\vtheta_n^H;\{\vz_m\}_{m=1}^M)$ 
			\end{itemize} 
			\item Take a descent step at the random point $s_n$:			\[\vtheta_{n+1}^H = \vtheta_n^H - \alpha_n^H \, \nabla_\vtheta L_H(\vtheta_n^H; s_n)\]
			\item Calculate the optimal control loss functional for the current mini-batch:
			\begin{itemize}
				\item $L_{u}(\vtheta_n^u;s_n) = -\tfrac{1}{M} \sum_{m=1}^M\left[ \Ls^{\vg(t_m, \vx_m; \vtheta_n^u)} f \left( t_m,\vx_m; \vtheta_n^H \right) + F\big( t_m,\vx_m,\vg(t_m,\vx_m;\vtheta_n^u) \big) \right]$				
			\end{itemize}
			\item Take a descent step at the random point $s_n$:
			\[\vtheta_{n+1}^u = \vtheta_n^u - \alpha_n^u \, \nabla_\vtheta L_u(\vtheta_n^u; s_n)\]
			\item Repeat steps (2)-(6) until $L_H$ reaches the tolerance $\varepsilon$ or for the number of iterations, $n=1,\ldots,N$.
		\end{enumerate}
		\vspace{0.5em}
		\begin{tabular}{r}			
			\hline\hline
			\hspace{0.95\textwidth}
		\end{tabular}
	}
	\vspace{-1em}
	\captionsetup{width=.9\linewidth}
	\caption{Modified Deep Galerkin Method for HJB Equations (DGM-PIA) algorithm. \label{fig:modifiedDGM}}
\end{figure}

\captionsetup[figure]{name={Fig}}

\begin{Remark}
A similar combination of PIA and neural networks was studied for a different class of PDE problems, namely semilinear Hamilton-Jacobi-Bellman-Isaacs (HJBI) boundary value problems, in \cite{reisinger2019neural_PIA}. Their algorithm, named \textit{inexact PIA}, considers a neural network approximation of the linear PDE (\ref{eq:linear_PDE}). The authors then analyze the convergence of this algorithm under this class of PDEs and prove its superlinear convergence.
\end{Remark}

In the subsequent sections we will consider various optimal control problems and numerically solve the associated HJB equations using the DGM-PIA algorithm and the direct application of the DGM algorithm to simplified HJB equations discussed earlier in this section. In each application we will present a very brief description of the problem; the reader is referred to the original papers or to the report by \cite{alaradi2018solving} for more complete descriptions and summaries.

\subsection{Merton Problem}\label{sec:merton}

In this section we apply the DGM-PIA algorithm to solve the Merton problem with exponential utility. Recall that in the Merton problem, an agent chooses the proportion of their wealth that they wish to invest in a risky asset and a risk-free asset through time. They seek to maximize the expected utility of terminal wealth at the end of their investment horizon; see \cite{merton1969lifetime} for the investment-consumption problem and \cite{merton1971optimum} for extensions in a number of directions. The HJB equation associated with this stochastic control problem is
{ \small
\begin{equation} \label{eqn:merton_unsimpHJB}
\begin{cases}
\partial_t H +  \underset{\pi \in \As}{\sup}  ~ \bigg\{ \big(\left( \pi (\mu -r) + rx \right) \partial_x + \tfrac{1}{2} \sigma^2 \pi^2 \partial_{xx} \big) H \bigg\}  = 0,
\\ 
H(T,x) = U(x),
\end{cases}
\end{equation}}%
where the model's parameters $\mu$, $\sigma$ and $r$ are the asset's drift and volatility and the risk-free rate, respectively. The proportion of wealth invested in the risky asset, $\pi$, is the agent's control. The state variable $x$ is the agent's wealth and $U$ is the agent's utility function. The HJB equation above could be simplified to
\begin{equation} \label{eqn:merton_simpHJB}
\begin{cases}
\partial_t H + rx \, \partial_x H - \displaystyle \frac{\lambda^2}{2} \frac{(\partial_x H)^2}{\partial_{xx} H}  = 0,
\\ 
H(T,x) = U(x),
\end{cases}
\end{equation}
where $\lambda = \frac{\mu - r}{\sigma}$ is the market price of risk; see Section 5.3 of \cite{cartea2015algorithmic}. 

If we assume an exponential utility function with risk preference parameter $\gamma$, that is $U(x) = -e^{-\gamma x}$, then the value function and the optimal control can be obtained in closed-form: 
\begin{subequations} \label{eqn:merton_soln}
\begin{align}
H(t,x)  &= - \exp \left[ {-x \gamma e^{r(T-t)} - \tfrac{\lambda^2}{2}  (T-t)	} \right] ,
\\
\pi^*(t,x) &=\yuri{-\frac{\lambda}{\gamma \sigma} \frac{\partial_x H(t,x)}{\partial_{xx} H(t,x)}} = \frac{\lambda}{\gamma \sigma} e^{-r(T-t)}.
\end{align}
\end{subequations}

\begin{figure}[h!]
\begin{minipage}{7in}
  \centering
  \raisebox{-0.5\height}{\includegraphics[width=0.5\linewidth, trim={0 1cm 0 1cm}, clip]{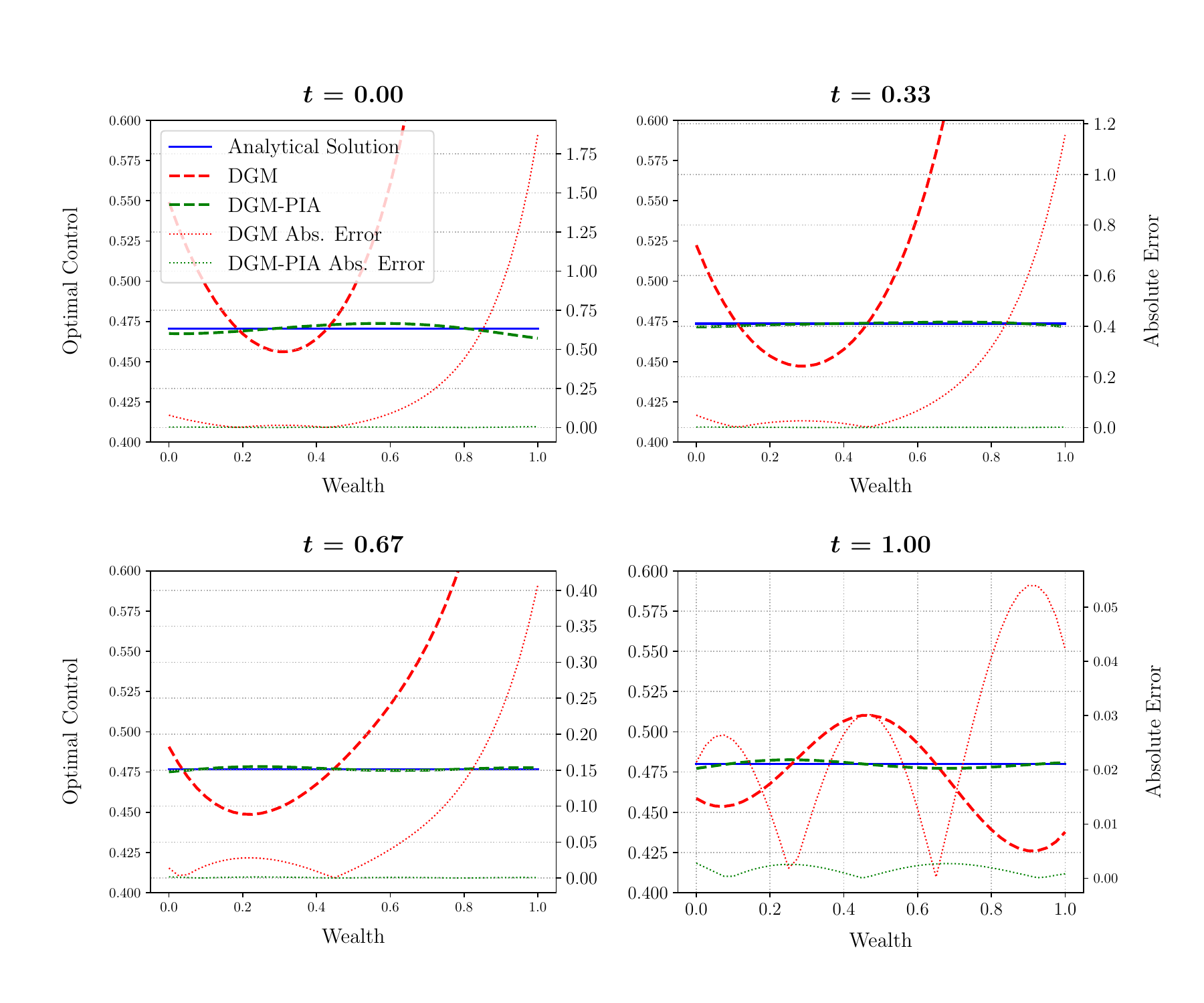}}
  \hspace*{-0.3in}
  \raisebox{-0.5\height}{\includegraphics[width=0.5\linewidth, trim={0 1cm 0 1cm}, clip]{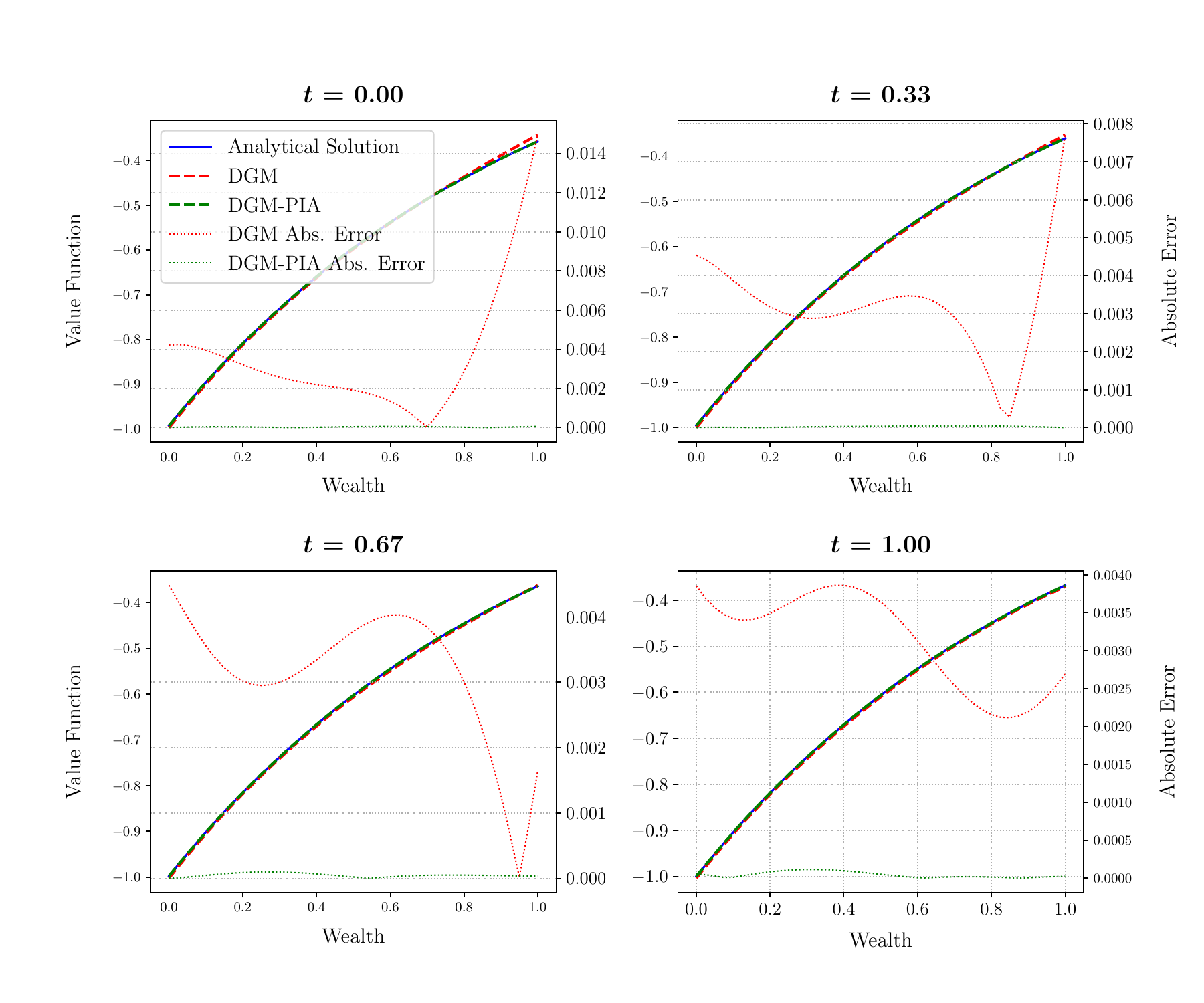}}
\end{minipage}
	\caption{Optimal control and value function of the Merton problem at different times \yuri{with their pointwise absolute error, their values shown in the right $y$-axis.}}
	\label{fig:MertonProblem}
\end{figure}

We apply DGM to \eqref{eqn:merton_simpHJB} and DGM-PIA to \eqref{eqn:merton_unsimpHJB} with parameters $r = 0.02$, $\mu = 0.05$, $\sigma = 0.25$, $\gamma = 1$ and $T = 1$. The approximated value function and optimal control compared to the analytical solution in \eqref{eqn:merton_soln} is given in Figure \ref{fig:MertonProblem}. Both algorithms ran for the same number of iterations.

\yuri{It is noteworthy that the DGM-PIA algorithm gives a more stable optimal control. Such behavior should be expected since the optimal control under the classical DGM is computed using the first and second derivatives of the approximated value function as in Equation \eqref{eqn:merton_soln}, magnifying the approximation errors. One should also notice that the DGM-PIA also delivers a better approximation for the value function. Additionally, we also present pointwise analytical and approximated values for both the value function and optimal control in Table \ref{tab:merton}.}

\begin{table}[h!]
\centering
\begin{tabular}{lccc}
\toprule
Wealth $x$  & Analytical  & DGM & DGM-PIA \\
\midrule
\multicolumn{4}{c}{Value Function} \\
\midrule
0.25  &   -7.6932e-01 & -7.7217e-01 & -7.6929e-01 \\
0.5   &   -5.9613e-01 & -5.9804e-01 & -5.9617e-01 \\
0.75  &   -4.6193e-01 & -4.6070e-01 & -4.6197e-01 \\
\toprule
\multicolumn{4}{c}{Optimal Control} \\
\midrule
0.25  &   4.7050e-01 & 4.5947e-01 & 4.6997e-01 \\
0.5   &   4.7050e-01 & 4.9530e-01 & 4.7353e-01 \\
0.75  &   4.7050e-01 & 7.7686e-01 & 4.7197e-01 \\
\bottomrule
\end{tabular}
\caption{\yuri{The pointwise analytical and approximated values at wealth values $x=0.25, 0.5$ and 0.75 and $t=0$.}} \label{tab:merton}
\end{table}

\subsection{Optimal Execution} \label{sec:optimalExecution}

HJB equations feature prominently in the algorithmic trading literature, such as in the classical work of \cite{almgren2001optimal} and more recently \cite{cartea2015optimal} and \cite{cartea2016incorporating} to name a few. In this section, we discuss a simple algorithmic trading problem with an investor that wishes to liquidate an inventory of shares but is subject to linear price impact and faces terminal and running inventory penalties. We omit a detailed discussion of the problem and refer the interested reader to Chapter 6 of \cite{cartea2015algorithmic} for additional details and for other optimal execution problems. For our purposes, we are interested in the HJB equation that arises in the context of this problem which is given by
\begin{equation} \label{eqn:optExec_unsimpHJB}
\begin{cases}
(\partial_t + \tfrac{1}{2} \sigma^2 \partial_{SS}) H - \phi q^2 + ~ \underset{\nu \in \As}{\sup} ~ \Big\{ \left(\nu(S - \kappa\nu)\partial_x - b\nu ~ \partial _S - \nu \partial_q \right) H \Big\} = 0,
\\
H(t,x,S,q) = x + Sq - \alpha q^2,
\end{cases}
\end{equation}
where $H$ is the unknown value function; $\nu$ is the agent's (liquidation) trading rate; the state variables $x,S$ and $q$ correspond to the investor's cash, the asset price and the investor's inventory; $\sigma$ is the asset's volatility; $k$ and $b$ are temporary and permanent price impact parameters; $\phi$ and $\alpha$ are running and terminal inventory penalty parameters. We start by carefully choosing the ansatz 
$$H(t,x,S,q) = x + qS + h(t,q).$$
The PDE (\ref{eqn:optExec_unsimpHJB}) becomes
\begin{equation} \label{eqn:optExec_unsimpHJB2}
\begin{cases}
\partial_t h - \phi q^2 + ~ \underset{\nu \in \As}{\sup} ~ \Big\{ \nu (-bq - \kappa\nu - \partial_q h )\Big\} = 0,
\\
h(t,q) = - \alpha q^2,
\end{cases}
\end{equation}
which could be simplified to
\begin{equation} \label{eqn:optExec_simpHJB2}
\begin{cases}
\partial_t h - \phi q^2 + \displaystyle \frac{(bq + \partial_q h)^2}{4\kappa} = 0,
\\
h(t,q) = - \alpha q^2.
\end{cases}
\end{equation}
We can then solve for the value function and optimal control:
\begin{subequations} \label{eqn:optExec_soln}
\begin{align}
h(t,q) &= \left( g(t) - \tfrac{b}{2} \right) q^2,
\\
\nu^*(t,q) &= \yuri{-\tfrac{1}{2k}(bq + \partial_q h(t,q)) = -\tfrac{1}{k}g(t)q},
\\
\mbox{where } \quad g(t) &= \sqrt{\kappa \phi} ~ \frac{1 + \zeta e^{2\gamma(T-t)}}{1 - \zeta e^{2\gamma(T-t)}} \, ,
\qquad 
\gamma = \sqrt{\frac{\phi}{\kappa}} \, ,
\qquad
\zeta = \frac{\alpha - \tfrac{1}{2}b + \sqrt{\kappa \phi}}{\alpha - \tfrac{1}{2}b - \sqrt{\kappa \phi}}  \,. \nonumber
\end{align}
\end{subequations}

\begin{figure}[H]
\begin{minipage}{7in}
  \centering
  \raisebox{-0.5\height}{\includegraphics[width=0.5\linewidth, trim={0 1cm 0 1cm}, clip]{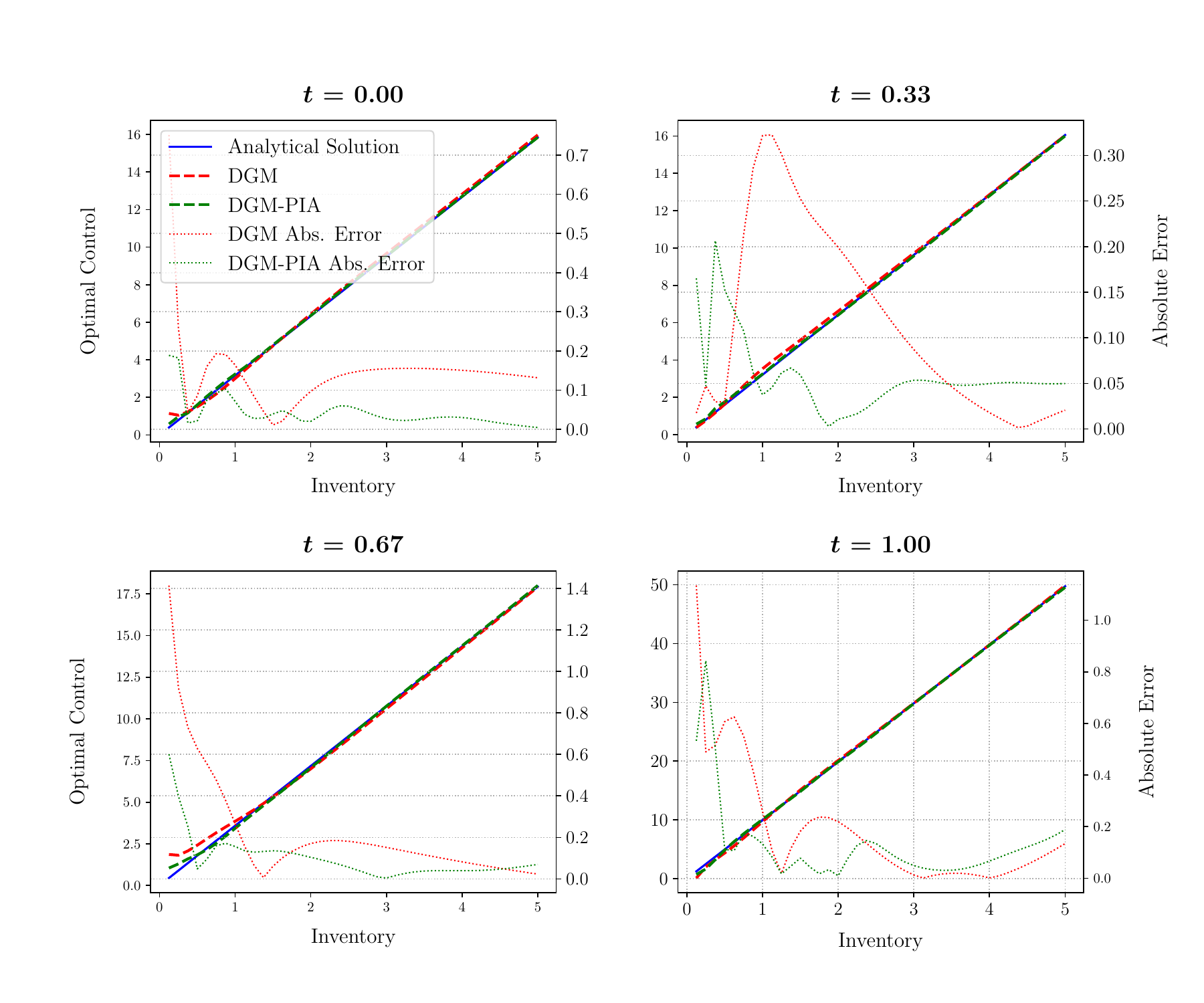}}
  \hspace*{-0.3in}
  \raisebox{-0.5\height}{\includegraphics[width=0.5\linewidth, trim={0 1cm 0 1cm}, clip]{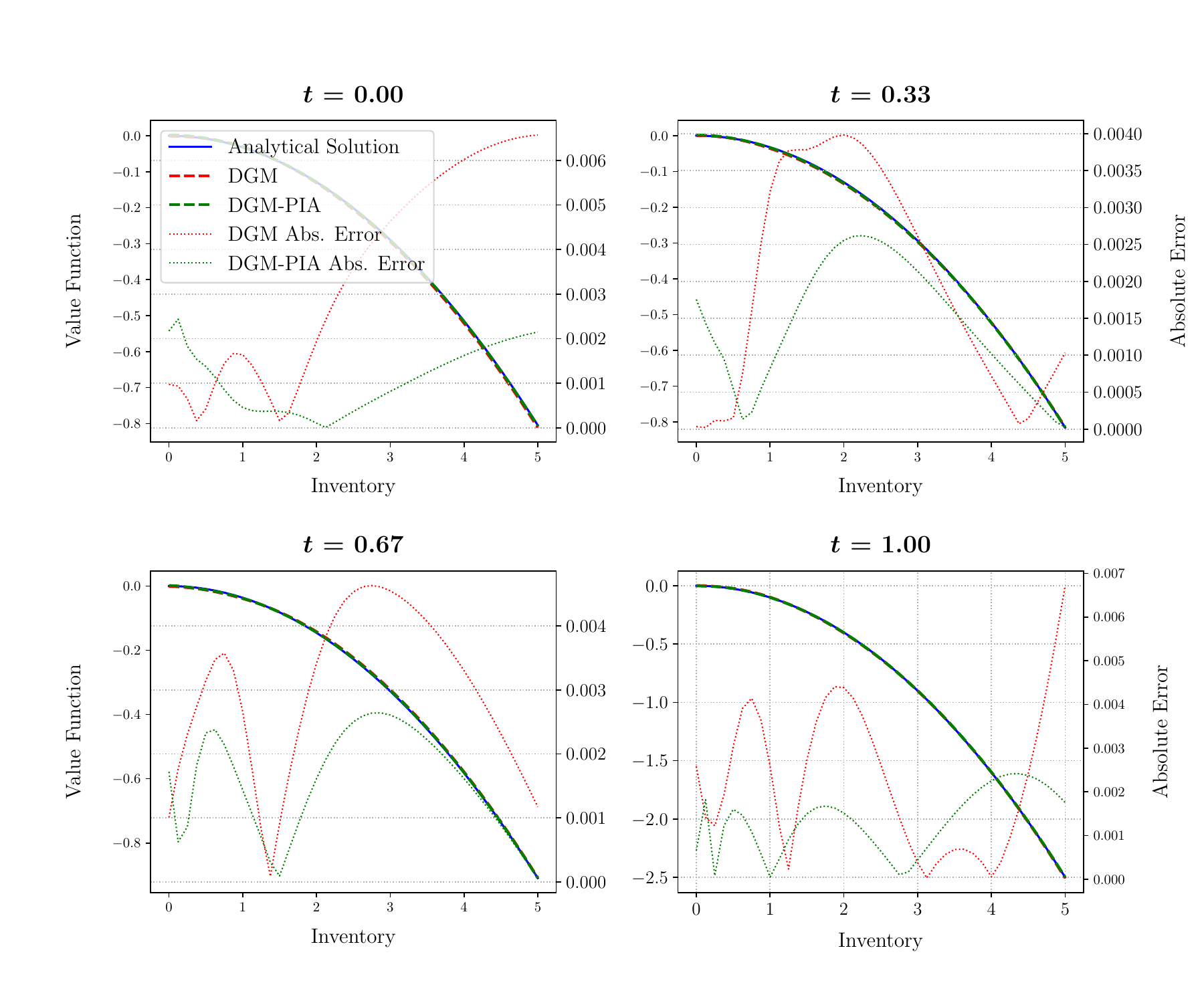}}
\end{minipage}
\caption{\yuri{Optimal control and value function of the Optimal Execution problem at different times with their pointwise absolute error, their values shown in the right $y$-axis.}}
	\label{fig:OptimalExecution}
\end{figure}

Now, we apply the DGM-PIA and DGM algorithms to the PDEs \eqref{eqn:optExec_unsimpHJB2} and \eqref{eqn:optExec_simpHJB2}, respectively, with parameters $k = 0.01$, 
$b = 0.001$, $\phi = 0.1$, $\alpha = 0.1$ and $T = 1$. The estimated value function and optimal control compared to the analytical solutions is given in Figure \ref{fig:OptimalExecution}. Both algorithms ran for the same number of iterations.

\yuri{As in the previous example, the modified DGM gives a more stable optimal control, mainly around $q = 0$, as demonstrated in Figure \ref{fig:OptimalExecution}. Notice however that the the improvement that the DGM-PIA brings, although clear, is less prominent when compared to the Merton problem presented in the section above. The reason is that the optimal control in this case depends only on the first derivative of the value function and in a more stable way; see Equation \eqref{eqn:optExec_soln}. Additionally, we also present pointwise analytical and approximated values for both the value function and optimal control in Table \ref{tab:execution}.}

\begin{table}[h!]
\centering
\begin{tabular}{lccc}
\toprule
Inventory $q$  & Analytical  & DGM & DGM-PIA \\
\midrule
\multicolumn{4}{c}{Value Function} \\
\midrule
1.25  &   -5.0284e-02 & -4.9236e-02 & -4.9915e-02 \\
2.5   &   -2.0113e-01 & -2.0471e-01 & -2.0150e-01 \\
3.75  &   -4.5255e-01 & -4.5830e-01 & -4.5399e-01 \\
\toprule
\multicolumn{4}{c}{Optimal Control} \\
\midrule
1.25  &   3.9602e+00 & 3.8764e+00 & 3.9876e+00 \\
2.5   &   7.9204e+00 & 8.0634e+00 & 7.9792e+00 \\
3.75  &   1.1881e+01 & 1.2034e+01 & 1.1912e+01 \\
\bottomrule
\end{tabular}
\caption{\yuri{The pointwise analytical and approximated values at inventory levels $x=1.25, 2.5$ and 3.75 and $t=0$.}} \label{tab:execution}
\end{table}

\subsection{Multidimensional Linear-Quadratic problem}

The third example we will study is the well-known linear-quadratic stochastic optimal control problem given by
\begin{align*}
F(t,\vx,\vu) &= \vx^\intercal Q \vx + \vu^\intercal R \vu, \quad G(\vx) = \vx^\intercal D \vx,\\
\
\Ls^{\vu} H(t,\vx) &= (A \vx + B\vu)^\intercal \nabla H(t,\vx)  + \frac{1}{2} \mbox{Tr}(CC^\intercal \nabla^2 H(t,\vx)),
\end{align*}
where $\nabla H$ is the gradient of $H$, $\nabla^2 H$ is the Hessian of $H$, the matrices $Q$, $R$ and $D$ are assumed to be symmetric and, additionally, $R$ is positive definite. 

The HJB in this case, in the unsimplified and simplified versions, are given by
\small{
\begin{align}\label{eq:LQ_pde}
\begin{cases}
\partial_t H + \underset{\vu \in \As}{\inf}\Big\{ \vx^\intercal Q \vx + \vu^\intercal R \vu + (A \vx + B\vu)^\intercal\nabla H + \frac{1}{2} \mbox{Tr}(CC^\intercal \nabla^2 H)  \Big\}  = 0,
\\
H(T,\vx) = \vx^\intercal D \vx.
\end{cases}
\end{align}}
and
\small{
\begin{equation*}
\begin{cases}
\partial_t H + \vx^\intercal Q \vx + (A \vx)^\intercal\nabla H + \frac{1}{2} (B^\intercal \nabla H )^\intercal R^{-1} (B^\intercal\nabla H) + \frac{1}{2} \mbox{Tr}(C^\intercal \nabla^2 H C)   = 0,
\\
H(T,\vx) = \vx^\intercal D \vx.
\end{cases}
\end{equation*}}
\normalsize

The closed-form solution for this non-linear, multidimensional PDE is given by
\begin{align*}
H(t,\vx) &= \vx^\intercal P(t) \vx + \int_t^T \mbox{Tr}(C^\intercal P(s) C) ds,\\
\vu^*(t,\vx) &= -\tfrac{1}{2}R^{-1} B^\intercal \nabla H(t,\vx) = -R^{-1} B^\intercal P(t) \vx,
\end{align*}
where $P$ solves the Ricatti ODE:
\begin{align*}
P'(t) &= P(t)BR^{-1}B^\intercal P(t) - A^\intercal P(t) - P(t)A - Q,\\
P(T) &= D.
\end{align*}
The particular case where all matrices are the $d$-dimensional identity, we find $H(t,\vx) = p(t) \|\vx\|^2 + d \, q(t)$ and $\vu^*(t,\vx) = -p(t)\vx$, where $p(t) = 1 + \sqrt{2} \tfrac{1 - e^{2\sqrt{2}(T-t)}}{1 + e^{2\sqrt{2}(T-t)}}$ and $q(t) = \int_t^T p(s)ds$. Note that $p(t) = -u^*_0(t,\mathbf{1})$ and $q(t) = H(t,\mathbf{0})$, and hence once we have approximations for $H$ and $\vu^*$, we can approximate $p$ and $q$. Therefore, we can study the accuracy of the approximation for the value function and optimal control.

\begin{figure}[H]
\centering
\includegraphics[width=\linewidth]{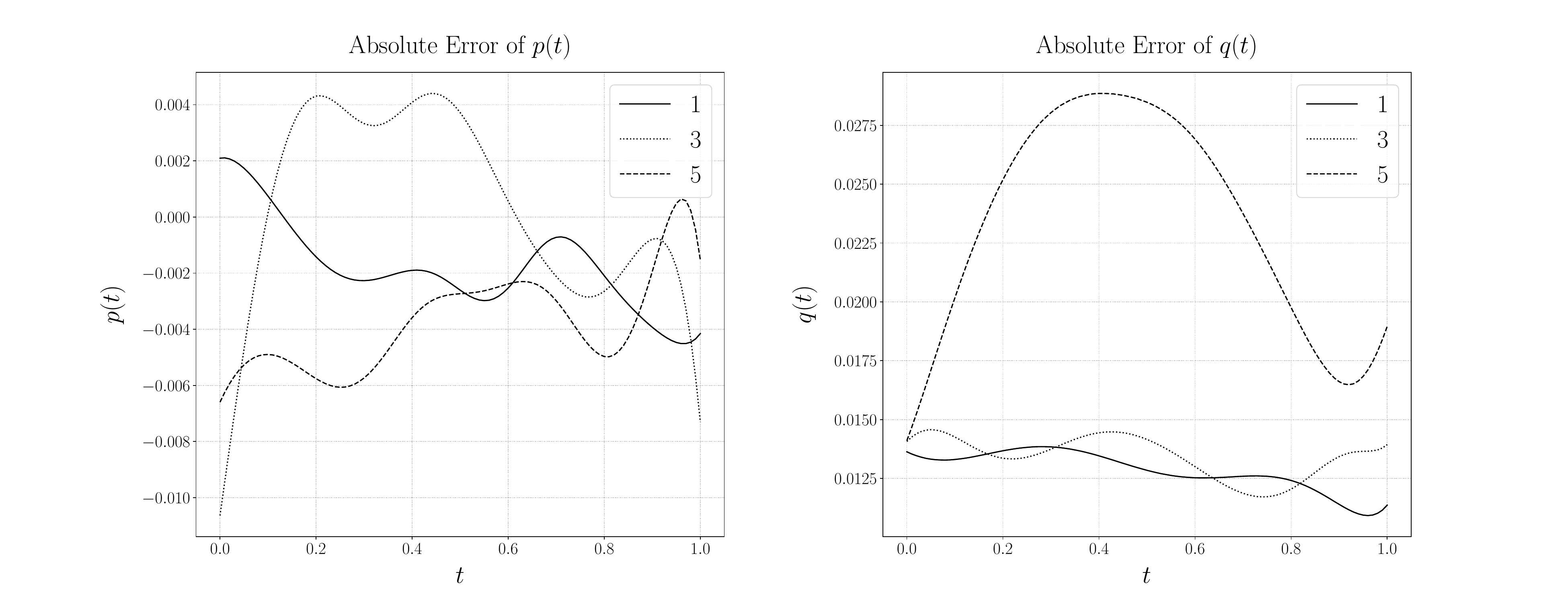}
\vspace{-1.2cm}
\captionsetup{width=.8\linewidth}
\caption{\yuri{Absolute error of the approximations for $p$ (left panel) and for $q$ (right panel)}} \label{fig:LQ}
\end{figure}

In Figure \ref{fig:LQ}, we show approximations of $p$ and $q$, respectively, for dimensions $d \in \{1,3,5\}$. We run the DGM-PIA algorithm for $N=50,000$ iterations. Additionally, we show some pointwise values for the different dimensions in Table \ref{tab:LQ}.

\begin{table}[h!]
\centering
\begin{tabular}{lcc}
\toprule
Dimension & DGM-PIA $p$  & DGM-PIA $q$\\
\midrule
1  &  1.8637 &  7.3304e-01\\
3  &  1.8573 &  7.3173e-01\\
5  &  1.8638 &  7.1740e-01\\
\bottomrule
\end{tabular}
\caption{\yuri{The pointwise analytical and approximated values at time $t=0.5$ of the functions $p$ and $q$. The analytical values are $p(0.5) = 1.8611$ and $q(0.5) = $7.4588e-01.}} \label{tab:LQ}
\end{table}

\section{Robustness}

\setcounter{figure}{0}

We have chosen some particular numerical values for the parameters that appeared in the PDEs in the examples so far. It is important to verify if the accuracy achieved has any dependence on this particular choice. Therefore, in order to assess the robustness of the proposed methods with respect to the numerical values of the parameters that appear in PDEs (\ref{eq:pde_u}) and (\ref{eq:hjb}), we perform the following robustness check. 

Consider a particular choice of the PDE such that a closed-form solution is available (for instance, we will exemplify with the Ornstein-Uhlenbeck studied in Section \ref{sec:OU} or the Merton problem in Section \ref{sec:merton}). Let us denote a possible parameter of this PDE by $\beta$, the true solution by $f(\cdot ; \beta)$ and the approximated solution by $\tilde{f}(\cdot ; \beta)$. We then measure the error for this particular parameter $\beta$ as the mean-squared error:
$$\mbox{MSE}(\beta) = \int_0^T \int_{\RR^d} (f(t,\vx;\beta) - \tilde{f}(t,\vx ; \beta))^2 d\vx \, dt$$
It is necessary to approximate the mean-squared error, so we consider the following approximation
\begin{align*}
\mbox{MSE}(\beta) \approx \frac{T (x_{\max} - x_{\min})^d}{n_t n_\vx} \sum_{i=1}^{n_t} \sum_{j=1}^{n_\vx} (f(t_i,\vx_j;\beta) - \tilde{f}(t_i,\vx_j ; \beta))^2,
\end{align*}
where $n_t$ and $n_\vx$ are the number of points in the time and space dimensions, respectively, and $\vx$ is taken uniformly (or equidistantly for small dimensions) in $[x_{\min}, x_{\max}]^d$. Moreover, we consider $N$ simulated values of the parameter $\beta$, $\{\beta\}_{i=1}^N$, where the draws are made from a uniform distribution $U[\beta_{\min}, \beta_{\max}]$. Finally, we visualize the results using a scatterplot of $\log($MSE$(\beta))$ against $\beta$.

Below we show the robustness of the methods proposed in Sections \ref{sec:FK_method} and \ref{sec:DGM_PIA} using the OU PDE and the Merton problem. For the OU PDE, we consider different values for $C$ (the volatility of $X$) sampled between 1 and 3, and for the Merton problem, we consider different values for $\sigma$ (the volatility of the risky asset) varying between 0.1 and 0.5.

\begin{figure}[h!]
	\centering
	\includegraphics[width=1.\linewidth, clip]{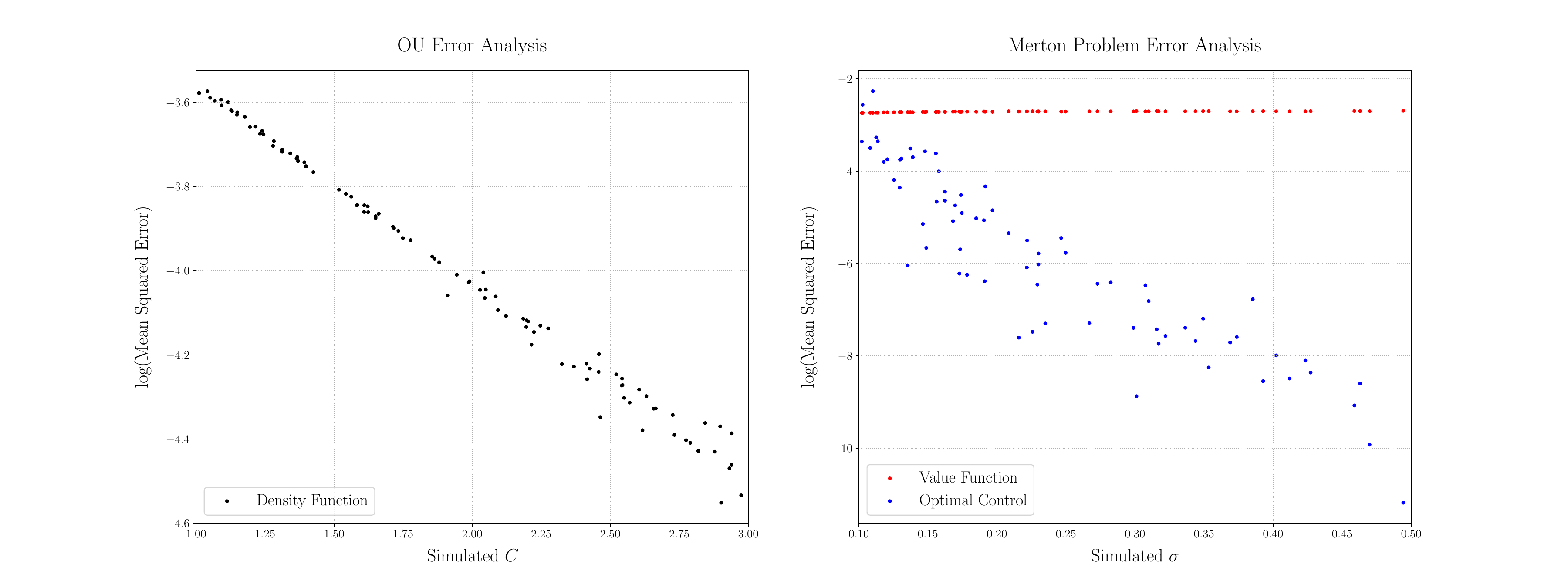}
	\caption{Log-mean-squared error for the OU problem with $B = 0.5$, $m=0$, $T=1$, $m_0 = 0$ and $C_0 = 1$ (left panel) and for the Merton problem optimal control and value function with $r = 0.02$, $\mu = 0.05$, $T=1$ and $\gamma = 1$ (right panel).}
\end{figure}

The MSE for the approximated density function of the OU problem stays small for different values of $C$. We observe that the error decreases the higher the value of $C$ and this might happen because the density becomes less concentrated around the mean, which might be easier for the algorithm to approximate.

The MSE for value function in the Merton problem stays almost constant in the log-scale for different values of $\sigma$. On the other hand, the MSE for the optimal control increases for small $\sigma$, which should be expected since the smaller the $\sigma$, the more complex the optimization step becomes, because the quadratic term is vanishing and then the unconstrained optimization becomes ill-defined. Nonetheless, the error remains small even for small $\sigma$.

\section{Systems of HJB Equations} \label{sec:HJBsystems}

\setcounter{figure}{0}

The next application we consider is based on the work of \cite{carmona2015systemic} on systemic risk which studies instability in a market where a number of banks are borrowing and lending with the central bank. Each player in this stochastic game aims to be at or around the average monetary reserve level across the economy.

We will focus on the system of HJB equations that characterize the optimal behavior of players and refer the interested reader to the original paper for additional details. The primal form of the HJB equation for agent $i \in \{1,...,n\}$ is
{ \small
\begin{equation} \label{eqn:sysRisk_unsimpHJB}
\begin{cases}
{\displaystyle
	\partial_t V^i + \underset{\alpha^i}{\inf} ~ \bigg\{ \sum_{j=1}^{N} \left[a(\overline{x} - x^j) + \alpha^j\right] \partial_j V^i + \frac{\sigma^2}{2} \sum_{j,k=1}^{N} \left( \rho^2 + \delta_{jk} (1-\rho^2) \right) \partial_{jk} V^i }
\\
{\displaystyle \hspace{5cm} + \frac{(\alpha^i)^2}{2} - q \alpha^i (\overline{x} - x^i) + \frac{\epsilon}{2}  \left(\overline{x} - x^i \right)^2 \bigg\} = 0, }
\\
V^i(T,\vx) = \frac{c}{2}  \left(\overline{x} - x^i \right)^2,
\end{cases}
\end{equation}
}%
where $\delta_{jk}=1_{\{j=k\}}$, $V^i$ is the value function for agent $i$; $\alpha^i$ is the agent's control which is the rate at which bank $i$ borrows from or lends to the central bank; $\vx = (x^1,...,x^n)$ are the state variables corresponding to the log-monetary reserves for each bank with $\overline{x}$ being the sample mean of this vector; $\sigma$ represents the volatility of the log-reserve and $\rho$ is the correlation between the independent Brownian motions that drive each log-rserve and a common noise, also modeled by a Brownian motion; $a$ is the mean reversion rate in log-reserves; $c,q$ and $\epsilon$ are preference parameters related to various running and terminal penalties. It is possible to arrive at a simplified system of HJB equations that do not contain an optimization step, given as follows:
\begin{align} \label{eqn:sysRisk_simpHJB}
\begin{cases}
{\displaystyle
	\partial_t V^i + \sum_{j=1}^{N} \left[(a+q)(\overline{x} - x^j) - \partial_j V^j \right] \partial_j V^i + \frac{\sigma^2}{2} \sum_{j,k=1}^{N} \left( \rho^2 + \delta_{jk} (1-\rho^2) \right) \partial_{jk} V^i }
\\
{ \displaystyle \hspace{5cm} + \tfrac{1}{2} (\epsilon - q^2) \left(\overline{x} - x^i \right)^2 + \tfrac{1}{2} \left( \partial_i V^i \right)^2 = 0, }
\\
{ \displaystyle V^i(T,\vx) = \tfrac{c}{2}  \left(\overline{x} - x^i \right)^2, \qquad \qquad  \qquad }
\end{cases}
\end{align}
for $i = 1,...,n$.

\begin{figure}[h!]
	\centering	
	\includegraphics[width=0.7\textwidth]{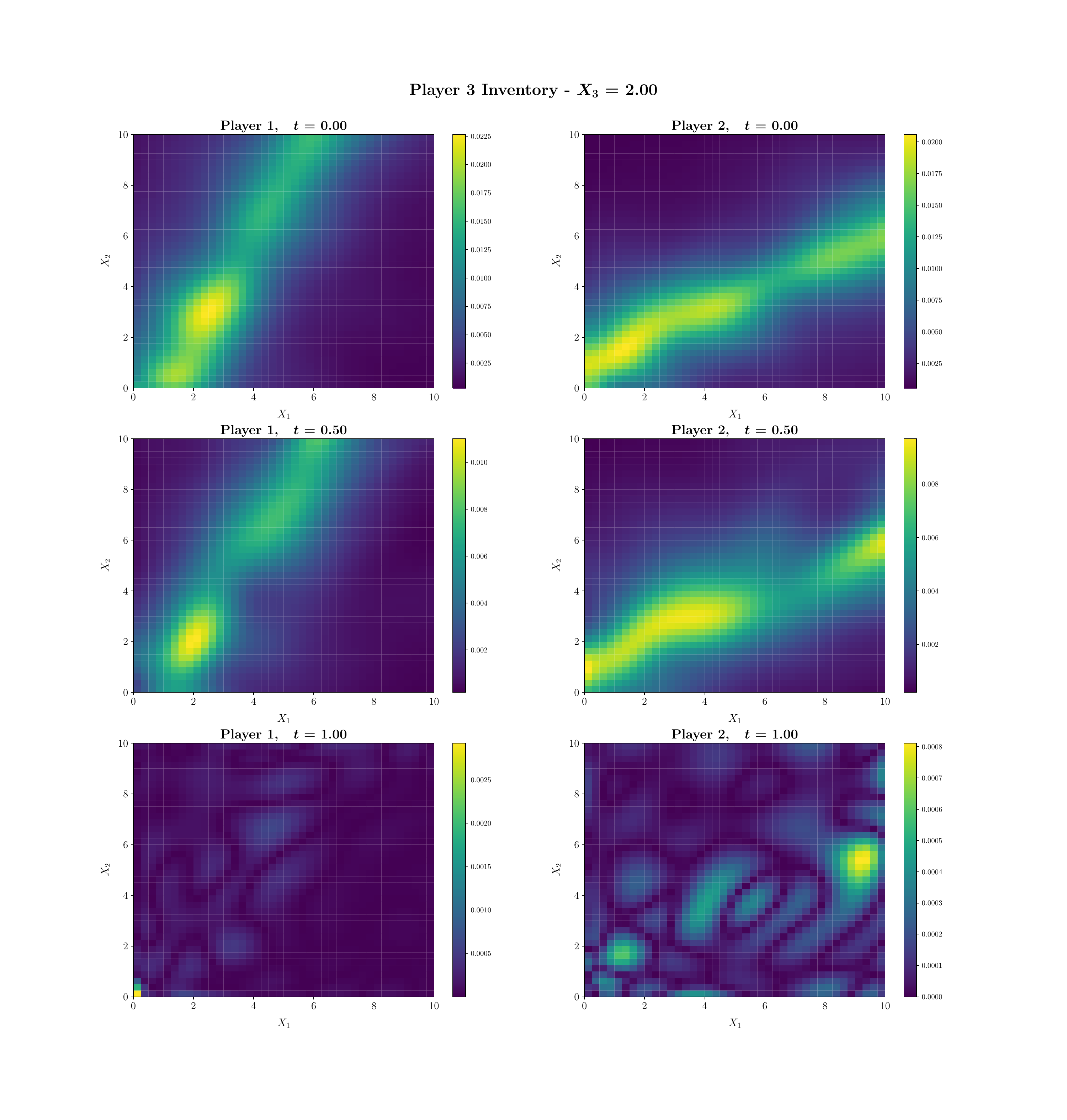}
	\captionsetup{width=0.9\linewidth}
	\caption{Relative errors using DGM in value function of players 1 and 2 at different times in the 3-player systemic risk problem when inventory of player 3 is equal to 2. We run the algorithm for 50,000 iterations.}
	\label{fig:systemicRisk_valueFunction_relErr1}
\end{figure}

Remarkably, this system of PDEs can be solved in closed-form to obtain the value function and the optimal control for each agent:
\begin{subequations} \label{eqn:sysRisk_soln}
\begin{align}
V^i(t,\vx) &= \frac{\eta(t)}{2} \left(\overline{x} - x^i \right)^2 + \mu(t), 
\\
\alpha_t^{i,*} &=  \bigg( q + \left(1 - \tfrac{1}{N} \right) \cdot \eta(t) \bigg) \left( \overline{X}_t - X^i_t \right),
\\
\mbox{where } ~~~  \eta(t) &= \frac{ -(\epsilon - q)^2 \left( e^{(\delta^+ - \delta^-)(T-t)} - 1 \right) - c \left( \delta^+ e^{(\delta^+ - \delta^-)(T-t)} - \delta^- \right)}{ \left( \delta^-e^{(\delta^+ - \delta^-)(T-t)} - \delta^+ \right) - c (1 - \tfrac{1}{N^2}) \left( e^{(\delta^+ - \delta^-)(T-t)} - 1 \right)}, \nonumber
\\ 
\mu(t) &= \tfrac{1}{2} \sigma^2 (1-\rho^2) \left( 1 - \tfrac{1}{N}  \right) \int_t^T \eta(s) ~ds, \nonumber
\\ 
\delta^{\pm} &= - (a+q) \pm \sqrt{R}, 
\qquad \qquad
R = (a+q)^2 + \Big( 1 - \tfrac{1}{N^2} \Big)(\epsilon - q^2) \nonumber
\end{align}
\end{subequations}
We apply the DGM algorithm to the system \eqref{eqn:sysRisk_unsimpHJB} for the three-player ($N = 3$) case with correlation $\rho = 0.5$, $\sigma = 0.2$, $a = 1$, $q = 1$, $\epsilon = 10$, $c = 1$, and $T = 1$ and compare the results with the analytical solution \eqref{eqn:sysRisk_soln}.  The DGM-PIA approach could be applied in this case with additional computational cost. The plot for the relative error for these results is given in Figure \ref{fig:systemicRisk_valueFunction_relErr1}.

\yuri{This is a very demanding PDE problem for any numerical method since it involves solving a system of nonlinear multidimensional PDEs. Under this setting, we are able to notice how flexible the DGM framework is achieving maximum relative error of around 2\% around the area of the function that is close to zero.}

\section{Mean Field Games} \label{sec:meanFieldGames}

\setcounter{figure}{0}

The final application we consider is based on the work of \cite{cardaliaguet2017mean} in the context of mean field games (MFGs), where the interest is in modeling the behavior of a large number of small interacting market participants. Building on the optimal execution problem, which was discussed in Section \ref{sec:optimalExecution} of this paper, \cite{cardaliaguet2017mean} propose extensions in a number of directions. First, traders are assumed to be part of a mean field game and the price of the underlying asset is impacted permanently, not only by the actions of an individual agent, but by the aggregate behavior of all agents acting in an optimal manner. In addition to this aggregate permanent impact, an individual trader faces the usual temporary impact effects of trading too quickly. The other extension is to allow for varying preferences among the traders in the economy. That is, traders may have different tolerance levels for the size of their inventories both throughout the investment horizon and at its end. Intuitively, this framework can be thought of as the agents attempting to ``trade optimally within the crowd.'' 

This application is of particular interest to us since it consists of both a system of HJB equations describing the optimal control problem of each individual agent along with a Fokker-Planck equation which governs the dynamics of the aggregate behavior of all agents. This forces us to use the techniques used in Sections \ref{sec:FKequations} in order to apply the DGM algorithm effectively. It is not straightforward to apply the DGM-PIA described in Section \ref{sec:HJBequations} to MFG PDEs because the Hamiltonian also appears in the Fokker-Planck equation.

Now, the HJB-Fokker-Planck system associated with the mean field game problem is:
{\small
\begin{equation} \label{eqn:MFG_simpHJB}
\begin{cases} 
\displaystyle  -\alpha \mu_t q =  \partial_t h - \phi q^2 +  \frac{(\partial_q h)^2}{4\kappa} = 0,
\\
h(T,q;\mu) =  -A q^2,
\\
~
\\
\displaystyle \partial_t u + \frac{1}{2\kappa}( h ~ \partial_q u - \partial_{qq} h) - \int_\RR \frac{e^{-u(t,q)}}{\int_\RR e^{-u(t,y)}~dy} ~ \partial_t u(t,q) ~dq = 0,
\\
\displaystyle u(0,q) = -\log ( \tilde{m}_0(q)),
\\
~
\\
\displaystyle \mu_t = \int_\RR  \nu(t,q) ~m(t,q)dq  \mbox{ and }
\displaystyle \nu(t,q)= \frac{\partial_q h(t,q)}{2\kappa}.
\end{cases}
\end{equation}
}%
The first two lines of system above correspond to the HJB equation associated with the optimal control problem faced by the representative agent. The variables in these equations are identical to the optimal execution discussed in Section \ref{sec:optimalExecution} with the addition of the state variable $\mu_t$ which corresponds to net sum of the trading rates of all agents and the parameter $\kappa$ which reflects the linear price sensitivity to this aggregate activity. The next three lines capture the evolution of the distribution of inventories across agents $m(t,q)$ and how this is driven by the net flow $\mu_t$ which in turn given by the aggregation of all agents' actions. 

The evolution of the density $m$, which begins at $m_0(q)$, through time is governed by a Fokker-Planck equation, and must also remain positive and integrate to unity. In order to apply the DGM algorithm we use the techniques discussed in Section \ref{sec:FKequations} to derive modified equations for the inventory density component that guarantees the numerical solution will be positive and integrate to 1. Using the same idea of exponentiating and normalizing used in Section  \ref{sec:FKequations}, we rewrite the density $m(t,q) = \frac{1}{c(t)} e^{-u(t,q)}$ where $c(t)$ is the required normalizing constant. Replacing the resulting PDE for the function $u$, the system for the MFG in the problem of \cite{cardaliaguet2017mean} becomes the second PDE shown in Equation (\ref{eqn:MFG_simpHJB}).

Moreover, notice that the Fokker-Planck PDE in principle requires us to know the formula of the Hamiltonian of the HJB equation. Therefore, it is not straightforward how the DGM-PIA algorithm presented in Section \ref{sec:HJBequations} should be adjusted in order to applied to the system above. We leave this for future work.

There is a closed-form solution for $h$. The form of the solution is fairly involved so we refer the interested reader to the details in \cite{cardaliaguet2017mean}. The paper also derives a closed-form expression for the expected inventory across agents through time, $E_t = \int_\RR q ~ m(t,q)dq$. We will use both the value function, optimal control and expected inventory to assess the accuracy of our numerical solutions.

\begin{figure}[h!]
\begin{minipage}{7in}
  \centering
  \raisebox{-0.5\height}{\includegraphics[width=0.5\linewidth, trim={0 1cm 0 1cm}, clip]{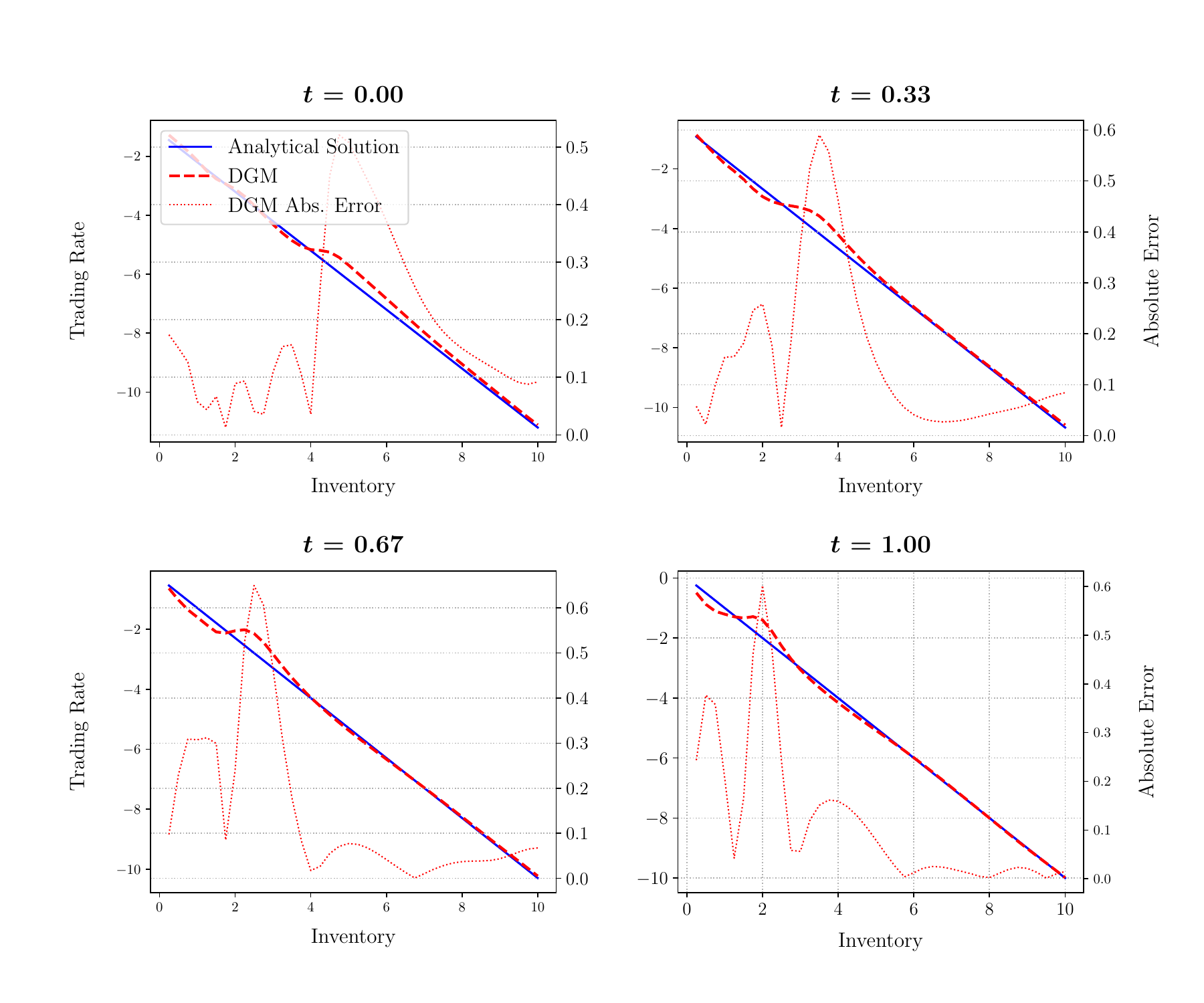}}
  \hspace*{-0.3in}
  \raisebox{-0.5\height}{\includegraphics[width=0.5\linewidth, trim={0 1cm 0 1cm}, clip]{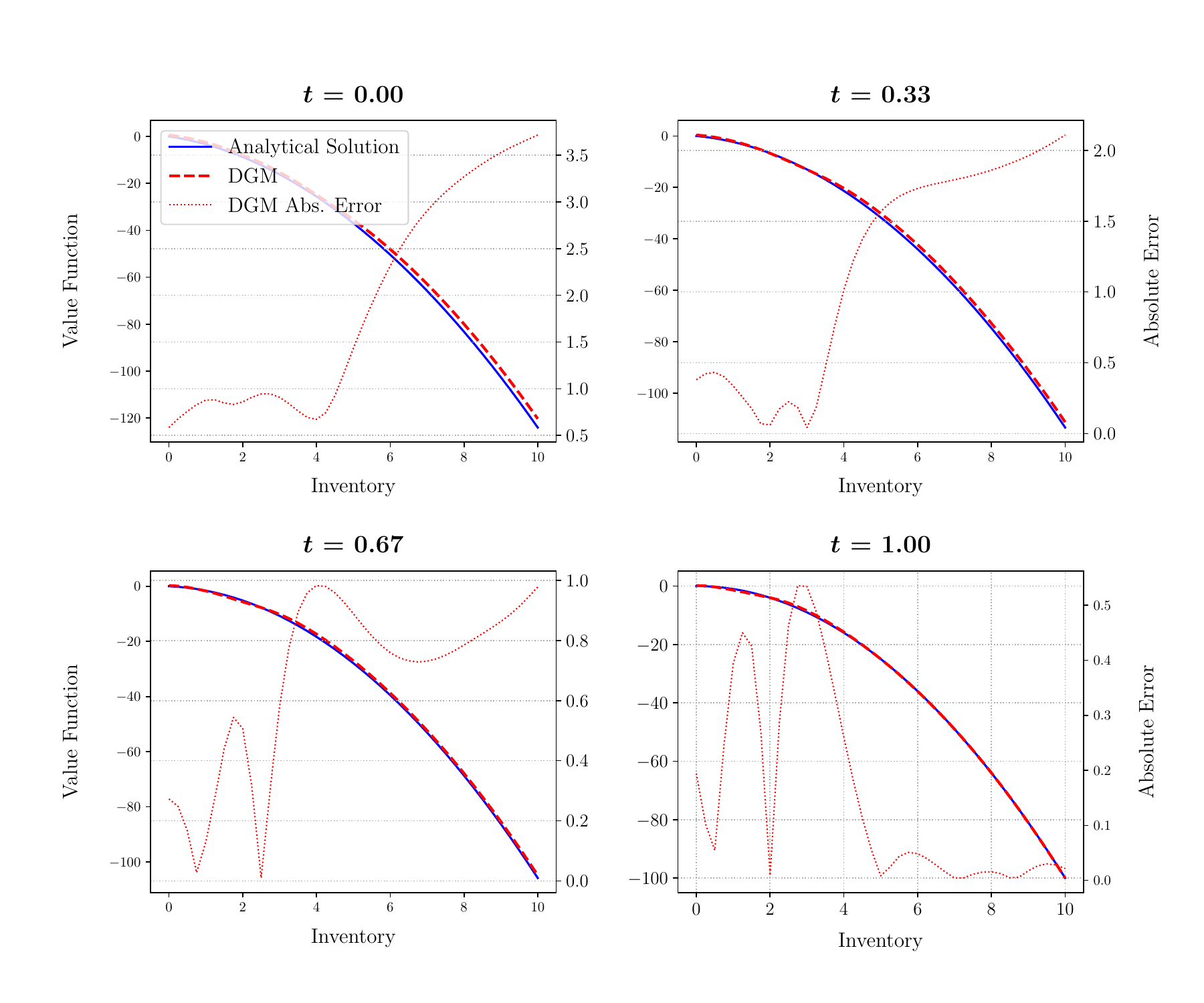}}
\end{minipage}
	\caption{\yuri{Optimal control and value function of the MFG problem at different times with their pointwise absolute error, their values shown in the right $y$-axis.}}
	\label{fig:MFG_value_control}
\end{figure}

\begin{figure}[h!]
\begin{minipage}{7in}
  \centering
  \raisebox{-0.5\height}{\includegraphics[width=0.5\linewidth, trim={0 1cm 0 1cm}, clip]{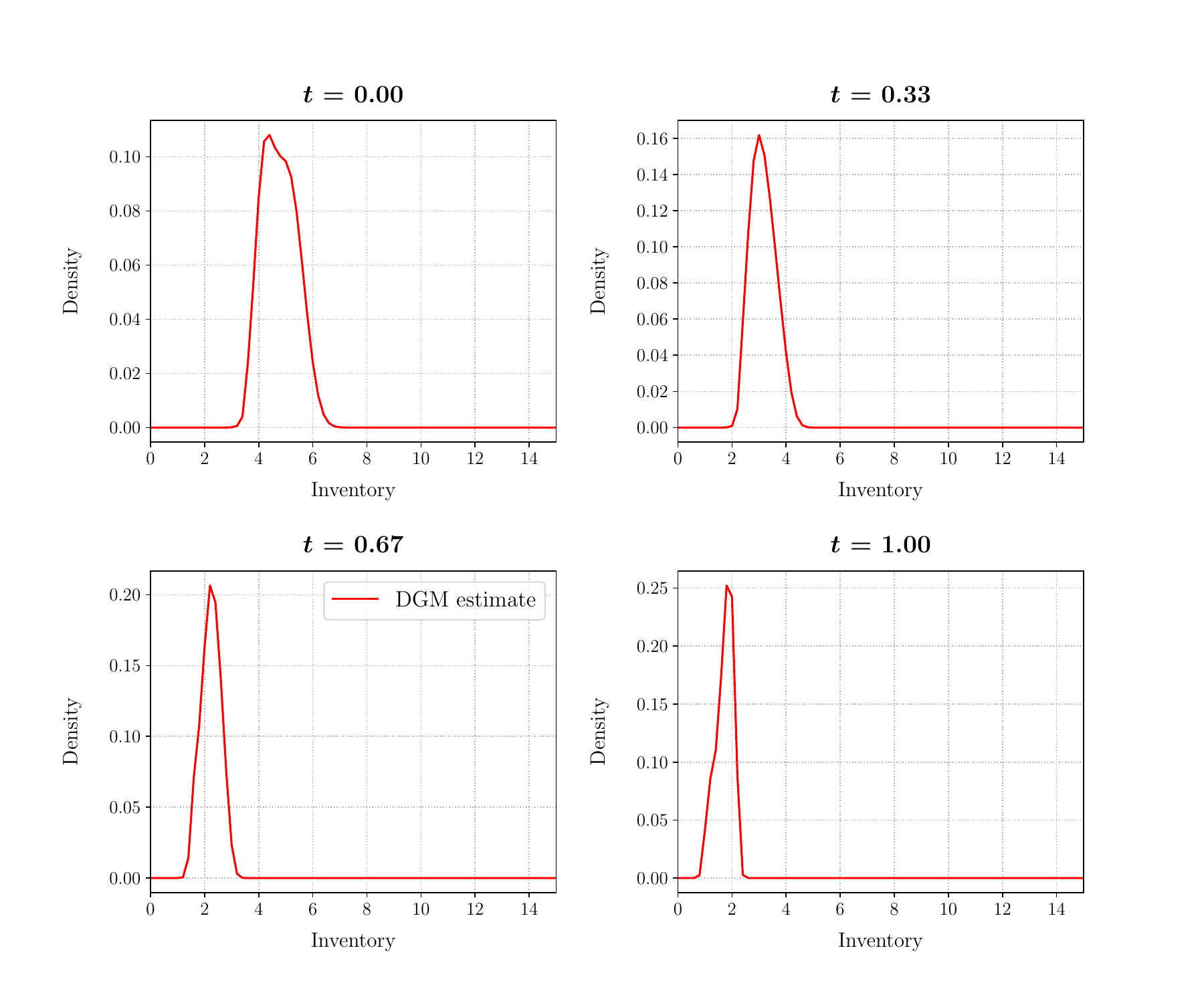}}
  \hspace*{0in}
  \raisebox{-0.5\height}{\includegraphics[width=0.3\linewidth, trim={0 1cm 0 1cm}, clip]{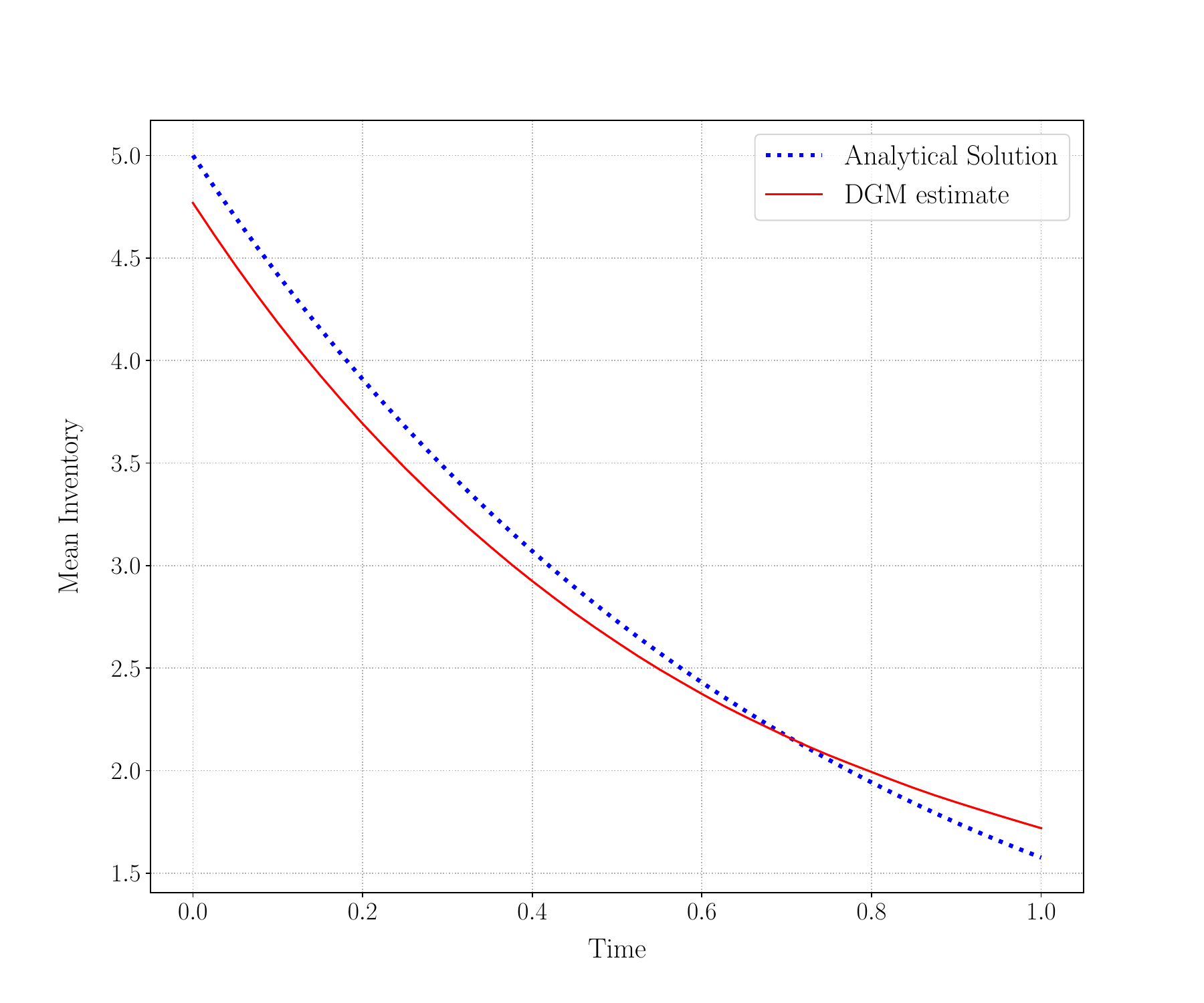}}
\end{minipage}
	\caption{{Distribution of agents' inventories at different times and the mean inventory across agents through time for the MFG problem .}}
	\label{fig:MFG_inv_expected}
\end{figure}

We apply the DGM algorithm to solve the system \eqref{eqn:MFG_simpHJB} with both integral terms being handled by importance sampling as in the Fokker-Planck equation with exponential transformation in Section \ref{sec:FKequations}. The system is solved numerically with parameters $A, \phi, \alpha, k = 1$ and with terminal time $T = 1$. The initial mass distribution is taken to be a normal distribution with mean $E_0 = 5$ and variance $0.25$. The value function, optimal control along with the expected values of the mass through time were compared with their respective analytical solutions (an analytical solution for the probability mass is not available; however the expected value of this distribution can be computed analytically). The resulting plots can be found in Figures \ref{fig:MFG_value_control} and \ref{fig:MFG_inv_expected}.

\yuri{One should notice that, although successful, the MFG example is a more challenging problem for the DGM algorithm, which is exemplified by the approximation for the optimal control in Figure \ref{fig:MFG_value_control}. As mentioned before, this example does not allow for analytical solution for the density $m$, but the behavior shown in Figure \ref{fig:MFG_inv_expected} is consistent with the theory that says that the agents are executing their shares and then holding less inventory as the times go forward up to the maturity $T$, also with smaller variance among the agents. The average behavior is known in closed form and presented in the left panel of Figure \ref{fig:MFG_inv_expected}.}

\section{Conclusions}
\label{sec:conclusions}

In this paper we presented an extension of the Deep Galerkin Method that uses ideas from policy improvement algorithms to solve HJB equations as well as PDEs involving constrained functions. The modified algorithm involves representing the value function and the optimal control as deep neural networks that are trained by taking alternating stochastic gradient descent steps. The algorithm is successfully applied to a number of optimal control problems that arise in financial contexts.

\appendix
\section{Proof of Theorem \ref{thm:reparam}} \label{sec:appendix_reparam}

Define
\[
v(t,\vx) = \frac{e^{-u(t,\vx)}}{c(t)} \qquad \text{where } ~c(t) = \int_{\RR^d} e^{-u(t,\vy)} \, d\vy
\]
we can find the derivatives of $v$ in terms of $u$ and $c$:
\begin{align*}
\partial_t v(t,\vx) &=  -v(t,\vx) \left(\partial_t u(t,\vx) +  \frac{c'(t)}{c(t)}\right),
\\
\partial_j v(t,\vx) &= - v(t,\vx) \partial_j u(t,\vx),
\\
\partial_{ij} v(t,\vx) &= v(t,\vx) (-\partial_{ij} u(t,\vx) + \partial_i u(t,\vx) \partial_j u(t,\vx)),
\end{align*}
where the time derivative of $c$ is given by 
\[
c'(t) = - \int_{\RR^d} \partial_t u(t,\vy) e^{-u(t,\vy)} \, d\vy \, .
\]
By the PDE (\ref{eq:pde_u}), we conclude that $v$ satisfies the Fokker-Planck PDE (\ref{eq:fokker_planck}). Moreover, the initial condition can be obtained by noticing that
\[
v(0,\vx) = \frac{e^{-u(0,\vx)}}{c(0)} = f(\vx) = p(0,\vx).
\]
Therefore, by uniqueness of solution of the Fokker-Planck PDE under Assumption \ref{assump:fokker_planck}, we conclude that $v = p$ and thus
\begin{equation*}
p(t,\vx) = \frac{e^{-u(t,\vx)}}{\int_{\RR^d} e^{-u(t,\vy)} \, d\vy}\, .
\end{equation*}
\begin{flushright}
\qed
\end{flushright}


\end{document}